\documentclass[11pt]{article}
\pdfoutput=1

\usepackage{times}
\usepackage{graphicx}
\usepackage{endfloat}
\usepackage{amssymb}
\usepackage{setspace}

\usepackage{JASA_manu}
\usepackage{amsmath}
\usepackage{graphicx}
\usepackage{longtable}
\usepackage{multirow}
\usepackage{amssymb}
\usepackage{subfig}
\usepackage{longtable} 
\usepackage{multicol}
\usepackage{comment}
\usepackage{natbib}
\usepackage{booktabs}

\usepackage{ragged}

\def\iid{\ensuremath{\stackrel{\text{\tiny iid}}{\sim}}}
\def\ind{\ensuremath{\stackrel{\text{\tiny ind}}{\sim}}}
\let\hat\widehat%
\let\tilde\widetilde%
\def\given{{\,|\,}}
\def\Pr{{\ensuremath{\mathbb P}}}
\def\Exp{{\ensuremath{\mathbb E}}}

\def\logit{\mbox{\rm logit}}

\newcommand{\blind}{0} % 1 for blinded, 0 otherwise

\begin{document}

\title{Perturbation Detection Through Modeling of Gene Expression on a Latent
Biological Pathway Network: A Bayesian hierarchical approach}
\if0\blind{
\author{%
Lisa M. Pham  \\
Luis Carvalho \\
Scott Schaus \\
Eric D. Kolaczyk \\
Boston University, Boston, MA 02215  \\ 
email: \texttt{lisamlpham@gmail.com}
}
}
\else{
\author{}
} \fi

\maketitle

%\newpage

\mbox{}
\vspace*{4in}
\begin{center}
\textbf{Author's Footnote:}
\end{center}
\if0\blind{
This research was supported, in part, by NIH grant GM078987. EK is partially
supported by AFOSR grant 12RSL042. LC is
supported by NSF grant DMS-1107067.  LP is supported by NIH/NIDDK grant R01 DK078616.

Mailing address: Program in Bioinformatics,
Boston University, 24 Cummington Street, Boston, MA 02215
(email: lisamlpham@gmail.com).
} \fi
\newpage
\begin{center}
\textbf{Abstract}
\end{center}

Cellular response to a perturbation is the result of a dynamic system of
biological variables linked in a complex network.  A major challenge in drug
and disease studies is identifying the key factors of a biological network
that are essential in determining the cell's fate.  

Here our goal is the identification of perturbed pathways from high-throughput
gene expression data. We develop a three-level hierarchical model, where (i)
the first level captures the relationship between gene expression and
biological pathways using confirmatory factor analysis, (ii) the second level
models the behavior within an underlying network of pathways induced by an
unknown perturbation using a conditional autoregressive model, and (iii) the
third level is a spike-and-slab prior on the perturbations. We then identify
perturbations through posterior-based variable selection. 

We illustrate our approach using gene transcription drug perturbation profiles
from the DREAM7 drug sensitivity predication challenge data set. Our proposed
method identified regulatory pathways that are known to play a causative role
and that were not readily resolved using gene set enrichment analysis or
exploratory factor models. Simulation results are presented assessing the
performance of this model relative to a network-free variant and its
robustness to inaccuracies in biological databases.

\vspace*{.3in}


%\newpage

\section{Introduction}

With the influx of high-throughput genomic data, understanding biological
mechanisms of action that cause changes in cellular homeostasis has become a
reachable challenge in the fields of bioinformatics, computational biology,
and statistics.  High-throughput measurement techniques such as
transcriptional profiling allow us to measure gene transcript levels across
thousands of genes simultaneously. However, analyzing individual gene
transcriptional profiles, cannot, by themselves, elucidate biological
mechanisms that are responsible for the changes observed in gene expression.
Rather, gene transcription provides only one facet of a multifaceted system of
biological variables that culminate to form a cellular response.

Mechanisms of action (MoA) that drive cellular dysregulation can arise in
several biological contexts. Chemotherapeutic compounds for instance, alter
very distinct mechanisms, such as changing the topology of DNA structure
induced by specific topoisomerase inhibitors (e.g. \citep{malik,nakada}), or
inhibiting cellular motility mechanisms caused by myosin II inhibitor
compounds  (e.g. \citep{allingham}). In another example, cancer metastasis is
also caused by aberrant pathways that disrupt normal cellular regulation,
resulting in cancer proliferation. The identification of such key mechanisms
is important as they can provide unique signatures not readily apparent by
directly analyzing gene expression without added structure or biological
context. Methods that leverage biological information by incorporating added
structure or biology can be used as a diagnostic tool in a clinical context.

Our primary goal in this paper is to identify drug targets and mechanisms of
action in drug perturbation experiments. That is, we aim to find pathways
significantly related to each drug's MoA, which can enhance both therapeutic
benefit and assessment of efficacy.  Moreover, understanding a drug's MoA
and biological pathway information can provide a better assessment of cellular
response to drugs, possibly providing a therapeutic profile for each drug
selection.

\subsection{The drug target problem: an inverse problem}

\begin{figure}[htbp]
\begin{center}
\includegraphics[width=\textwidth]{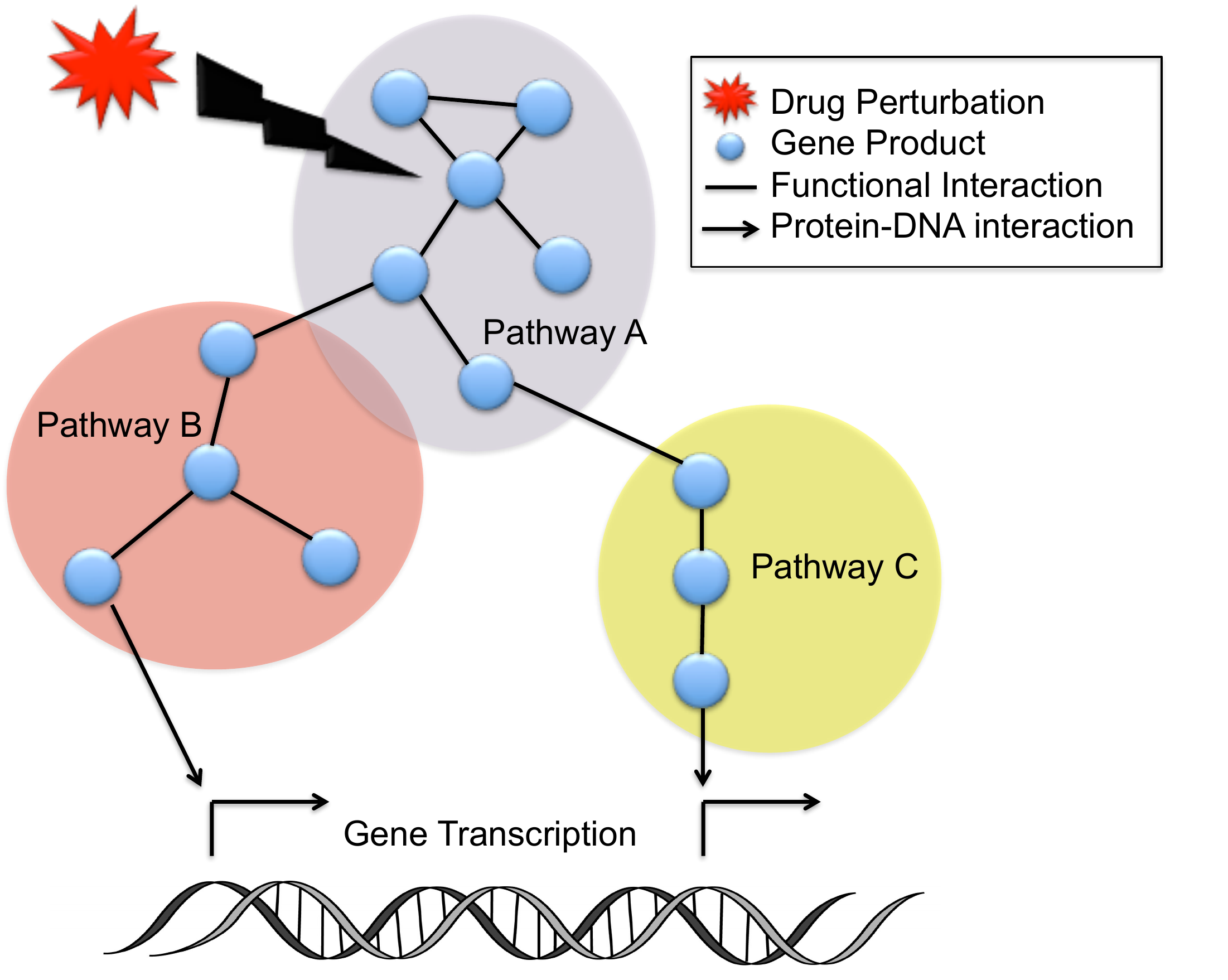}
\caption{A representation of a drug perturbation to a pathway A and its
downstream effects on associated pathways and gene expression.}
\label{MoA_illustration}
\end{center}
\end{figure}

A cellular response is the result of a culmination of interactions between
genes/proteins. A single perturbed pathway, for instance, can cause a rippling
effect across a global network of interactions, leaving behind a cascade of
transcriptional dysregulation across the genome. For instance,
Fig.~\ref{MoA_illustration} is a schematic illustration of how a drug perturbs
a system of interacting pathways.  In Fig.~\ref{MoA_illustration}, the
inhibition/activation of a protein (s) in a single pathway consequently alters
several downstream pathways, leading to changes in gene expression.  To better
understand the primary targets of a perturbation source, we propose a method
that formally filters out these regulatory dependencies between biological
factors (pathways) to uncover the primary
target underlying a perturbation response.

From a mathematical perspective, the nature of the problem we address in this
paper is not unlike that of `deconvolution' in image processing, a similarity
that has been noted by others in this area (e.g., `drug target deconvolution'
~\citep{terstappen}). In the image processing version of this problem, an
image, say $f$, is of interest but one has available only blurred and noisy
measurements, say $y=Kf+e$.  While denoising $y$ can be relatively
straightforward, it only leaves one with an estimate of the blurred image,
$Kf$.  In order to recover $f$ itself, the effect of the blurring operator $K$
must be inverted.  However, even in the ideal case where $K$ is known this
inversion can be ill-posed and the recovery of $f$ can be severely degraded by
the corresponding inflation of the noise $e$.  When $K$ is unknown or only
partially known, as is analogous to what we face in the drug target prediction
problem, the degradation can be arbitrarily worse.

\subsection{Identifying Pathway Targets in the DREAM 7 Drug Sensitivity
Prediction Challenge Data Set}

For our purposes of target pathway identification in drug perturbation
experiments, we explore the NCI DREAM7 drug sensitivity prediction
challenge dataset \citep{bansal} which is a part of the Dialogue for
Reverse Engineering Assessments and Methods (DREAM) challenge series
\citep{dream5, dream4}. To assess the performance of our method, we
focus our attention on the DREAM7 drug sub-challenge~2 dataset
\citep{bansal} which consists of microarray gene expression profiles from the
LY3 cancer cell line. Exactly~14 drugs were tested at different concentrations
and durations, and were compared to their mock control counterparts.  These
high quality, methodical, and carefully designed experiments serve well in
testing methods that are designed to predict drug mode of actions because
their cellular effects have been well studied, spanning a variety of
mechanisms from DNA-damaging agents (e.g.\ etoposide \citep{nakada}) or
cellular motility inhibitors (e.g.\ blebbistatin \citep{allingham}) to
compounds that disrupt regulatory signaling mechanisms (e.g geldanamycin
\citep{neckers,grenert}).

Differential gene expression analyses and other gene enrichment methods may
provide insight into dysregulated genes or gene sets (e.g.\ biological
pathways)  resulting from a drug perturbation propagating through a system of
interacting genes or proteins.  However, identifying the primary source of
perturbation that can explain the global variation in gene expression is often
difficult to discern from differential gene expression alone.  For
instance, DNA damaging agents that induce cell cycle arrest initiate a series
of biological processes such as cell death pathways (apoptosis), protein
degradation pathways (e.g. RNA degradation, ubiquitin mediated proteolysis),
and possibly DNA-repair pathways.   As a result, genes associated with these
downstream pathways may be upregulated and consequently, identified by
differential gene analyses such as gene set enrichment analysis
\citep[GSEA,][]{gsea}.  Rather than detecting the residual effects of such a
perturbation, we aim to identify upstream pathways positioned to cause changes
in gene expression.  In fact, in the case of DNA damaging agents such as the
drug camptothecin, we identified P53 signaling in the DREAM7 dataset while
GSEA has not (see Section~\ref{sec:ic20-12} for details); P53 signaling may be
causally linked to cell cycle arrest induced by DNA damage \citep{jaks,gupta,wang}.

Moreover, comparing drug profiles from two different exposure times, we show that
certain drugs are more sensitive on the LY3 cancer cell line than others.  We
also identified drug-induced pathways that were consistently identified across
varying conditions.  Lastly, we found that drugs having similar mechanisms
(e.g. DNA damaging agents) clustered together using profiles generated by our
method.

%\subsection{Model Overview}
%In this paper, we aim to uncover these biological MoAs underlying gene
%expression data.  We approach this as a task in statistical
%modelling and inference using a biological network of pathways, where each
%node represents a canonical pathway.  We develop a two-level statistical
%model, where (i) the first level is a confirmatory factor analysis (CFA) model
%\citep{sem} that captures the relationships between gene expressions and
%biological pathways, and (ii) the second level is a conditional autoregressive
%(CAR) model \citep{cressie,oliveira} that models the behavior of a pathway
%within an underlying network of pathways induced by an unknown perturbation.
%Detection of perturbed pathways is accomplished through statistical inference
%on latent variables representing perturbation targets, using principles of
%classical Bayesian analysis [FIXME:  this sentence is too vague]. 

\subsection{Organization of this paper}

In Section~\ref{prior_work}, we discuss related work.  In
Section~\ref{sec:model_framework}, we describe the hierarchical model in
detail including priors and model identification constraints. In
Section~\ref{sec:sampling}, we outline the aims of posterior inference and the
steps to our sampler.  We assess the performance of our model compared to an
exploratory factor analysis (EFA) model using simulated data sets in
Section~\ref{sec:simulation}.  We discuss our results after applying our
method to a drug perturbation dataset (DREAM7 \citep{bansal}) and compare our method
against an EFA model and gene set enrichment analysis (GSEA) \citep{gsea} in
Section~\ref{dream7}.  Lastly, in Section~\ref{sec:discussion}, we summarize
our method as well as our results.

\section{Prior and Related Work}
\label{prior_work}
Currently, there are roughly three ways the task of identifying disease or
drug targets have been approached.  In the simplest approach, statistical
methods are used to find differential expression between a binary phenotype
\citep{Dudoit,Storey}. The goal with this approach is to identify individuals
genes whose expression levels are associated with a trait relative to some
control conditions.  These methods are rooted in multiple testing and usually
control either the family-wise error rate \citep{Dudoit} or the false
discovery rate \citep{Dudoit,Storey}. Although relatively straightforward,
these methods have two main drawbacks: they can only identify downstream
transcriptional effects since the goal is to detect differential expression
and not possible sources of perturbations; and they operate on a gene-by-gene
basis and so the phenotypes for each gene are compared in isolation instead of
in coordination across gene sets.

The second approach, however, uses biological information to group genes
with the goal of capturing coordinated expression changes within a gene set
that might not have been detected if the analysis were conducted in individual
genes. The aim, therefore, is to look for statistical significance in
differential expression across entire gene sets.  These gene sets represent
a system's level view of the cell such as cellular pathways or transcription
factor targets. The most simple and straightforward method to test if a gene
set contains a significant amount of transcriptionally altered genes is a
hypergeometric test.  This method and other similarly straightforward methods
are reviewed in \citep{Khatri,Rivals}. \citet{gsea} developed a method called
gene set enrichment analysis (GSEA) with the same goal, but this method does
not require a user to choose a threshold to separate differentially expressed
genes from non-differentially expressed genes.  Instead, GSEA measures how
well a gene set clusters at the top or bottom of a \emph{ranked} list of genes
(e.g.\ ranked using fold ratios or p-values from two sample tests when
comparing binary phenotypes).
%GSEA scores predefined gene sets according to an algorithm that examines how
%well the genes within the gene set will cluster at the top or bottom of a list
%of genes ranked by differential gene expression between a binary phenotype.
Extensions of GSEA and other threshold-free methods are described in
\citep{Jiang,Nam}. GSEA is the most widely used gene set analysis tool to
date, however it does not use any gene network information.

The third and final approach is network-based and uses information from
cellular regulation or gene-protein interactions to provide a better
understanding of the molecular mechanisms underlying a response. On an
individual protein-gene level, many supervised methods have been proposed to
predict interactions between drugs and proteins. For instance, using existing
information of known protein-drug interactions, machine learning methods
\citep{faulon,yamanishi} have been used to predict interactions of individual
proteins and drugs using a combination of protein sequence and chemical data.
Network based approaches that identify gene sets over individual genes or
proteins have also been proposed. \citet{Gu}, for instance, use a
two step method where in the first step they identify a set of differentially
expressed (DE) genes that are then mapped to a biological network; in the
second step, a clustering method is used to identify genes that form a
connected subgraph of DE genes. Other techniques, including pathway-express
\citep{pathwayexpress} and gene network enrichment analysis \citep{gnea}
have also used network topology. Pathway-express \citep{pathwayexpress} scores
genes in a network using differential expression such that differentially
expressed genes that are connected are scored higher than those that are not
connected.  Pre-defined gene sets are then represented using the scores of its
genes in the network.
Gene network enrichment analysis \citep{gnea} first identifies a connected
subnetwork of differentially expressed genes of a global protein-protein
interaction network, and then identifies pre-defined gene sets, such as
biological pathways, that are enriched with respect to the genes in a
differentially expressed subnetwork.

Although these increasingly complex approaches serve well in identifying genes
or gene sets that are differentially expressed, they are not designed to
identify which genes or gene sets, of the collection of genes or gene sets
identified by their method, are primary targets of a specific response, and
thus they do not prescribe a mechanism of action.
To this end, it is essential to explicitly model the cascading effects of a
perturbation by considering biological relationships in a gene network.
Our proposed method is such a model-based approach: a factor model, called
confirmatory factor analysis (CFA) with a conditional autoregressive (CAR)
component designed to identify primary targets underlying gene expression.
Spatial conditional autoregressive models have recently been explored in
biology, specifically with functional networks (e.g. \citep{wei}). The
motivation behind such models stems from the principle that genes that work
together (i.e.\ genes that are functionally related) are usually
co-expressed/inhibited, co-\@(de)\@regulated, or co-\@(de)\@activated.

Factor analysis models have also been used extensively in biology
\citep{west_2006,west_2008,lucas,ma}.  \citet{ma} use a Bayesian
factor analysis model to identify drug-pathway interactions by analyzing
paired gene expression and drug sensitivity data and their relationship to
common pathways, represented as latent factors.  In their model all factors
(pathways), however, are independent.  These network-free models are referred
to as exploratory factor analysis (EFA) models.  West et al.
\citep{west_2006,west_2008} also uses exploratory factor analysis models but
with an aim to identify biomarkers using gene expression data.  However, in
\citep{west_2006,west_2008} the factors are not structurally informed by known
biological gene sets and instead are completely data-driven. Lastly,
EFA or traditional CFA models assume a zero prior mean for the latent factors.
Conversely, our proposed method is a novel approach to perturbation target
identification on a latent scale which give rise to non-trivial prior means on
the latent factors. Together, our proposed method is a synthesis of several
key principles from both confirmatory factor models and conditional
autoregressive models that allow us to explain the variation observed in
transcriptional data as a function of perturbations to a latent network of
biological pathways.

Our proposed model was first motivated by the work of \citet{bernardo_2005},
where a mathematical model adapted from Hill-type transcription kinetics
\citep{liao_2003} was used to describe the relationships among gene
transcripts in a cell. The model captures both internal regulatory influences
among $p$ gene transcripts and effects due to external perturbation effects:
\begin{equation}
\label{bernardo}
\log_{10} \Bigg( \frac{\nu_i}{\nu_{ib}} \Bigg) =
\log_{10}  \Bigg( \frac{\mu_i}{\mu_{ib}} \Bigg) +
\sum_{j=1}^p \eta_{ij} \log_{10} \Bigg( \frac{\nu_j}{\nu_{jb}} \Bigg),
\end{equation}
where $\nu_i$, for $i = 1, \ldots, p$, represents the expression level of
transcript $i$, $\mu_i$ represents the direct influence of the perturbation on
transcript $i$, $\nu_{ib}$ and $\mu_{ib}$ represent a set of baseline values
of $\nu_i$ and $\mu_i$, respectively, and $\eta_{ij}$ represents the influence
of transcript $j$ on transcript $i$.
%With microarray data, the ratios with respect to baseline values are replaced
%by RMA-normalized expression values.
This can be re-expressed as 
\begin{equation}
\label{bernardo_2}
{\bf \psi} = B {\bf \psi} + \phi
\end{equation}
where {\bf $\psi$} = $( \psi_1, \ldots, \psi_p )'$ with $\psi_i =
\log_{10}(\nu_i / \nu_{ib})$, and {\bf $\phi$} = $( \phi_1, \ldots, \phi_p)'$
with $\phi_i = [1/(1-\eta_{ii})]\log_{10}(\mu_i/\mu_{ib})$, a scaled version of
the log-relative direct influence of the perturbation.  Here, $B$ is a
$p\times p$ matrix representing the network interaction effects among gene
transcripts with $B_{ii} = 0$ and, for $i \neq j$, $B_{ij} = \eta_{ij}/
(1-\eta_{ii})$.

The work in \citep{cosgrove_2008} is a statistical extension of the
mathematical model in Eq.~\eqref{bernardo_2} in the form of 
\begin{equation}
\label{cosgrove}
{\bf y} = B {\bf y} + \phi + e
\end{equation}
where $y$ is now the $p\times 1$ vector of measures of $\psi$.  Here, each
measured expression level $y_i$ is potentially influenced by other
transcripts, where this influence is captured through $B$.  The perturbations
are additive components represented by $\phi$. 

Here, we further extend this model to factor analysis models which are
frequently used to analyze high-dimensional gene expression data
\citep{ma,lucas,west_2008,west_2006}.  Here, we assume that gene expression is
linked to an overall factor defined by biological pathways in the cell.  Our
proposed model can be described as a hierarchical model where the first (gene)
level is a regression of gene expression on biological pathways, and the
second (pathway) level is an auto-regressive model that models the behavior of
\emph{a system of interacting biological pathways} underlying external
perturbation effects.  We describe the proposed model in
Section~\ref{sec:model_framework}.

\section{Model Framework}
\label{sec:model_framework}

We developed a statistical hierarchical model to uncover perturbations to
biological pathways using high-throughput measurements of gene expression on
individual genes.  The data $Y = \{Y_e,Y_c\}$ is a $p \times n$ matrix of
transcriptional profiles of $p$ genes across $n$ varying conditions, and
consists of a set of cases, $Y_e$, and controls, $Y_c$.  To avoid using an
intercept in our model, we center the data with respect to the mean of the
control group, $Y_c$. Our goal is to identify latent factors (e.g.\ biological
pathways) that can explain the variation observed in the experimental data,
$Y_e$, relative to the control conditions, $Y_c$.

Our model is built on three hierarchical components:
\begin{align}
y_{ki} \given \Lambda, \omega_i, \psi_k &\ind
N(\Lambda_k \omega_{i}, \psi_k) \label{cfa_model} \\
\omega_{ji} \given \omega_{[-j]i}, \rho_{ji}, \sigma^2 &\ind
N\Bigg( \sum_{j'=1, j' \neq j}^{q} B_{jj'} \omega_{j'i}
+ \rho_{ji}, \sigma^2\Bigg) \label{car_model} \\
\rho_{ji} \given \theta_{ji}, \tau^2 &\ind
N(0, \theta_{ji} \tau^2 + (1 - \theta_{ji}) v_0 \tau^2)
\label{vs_model}
\end{align}
for $i = 1, \ldots, n$, $k = 1, \ldots, p$, $j =  1, \ldots, q$, where 
$n$, $p$, and $q$ represent the number of samples, genes, and pathways,
respectively, and
	\begin{itemize}
  \item $ y_{ki} $ is the observed gene expression level of gene $k$ in sample
  $i$.
  \item $ \omega_{i} $ is a $q \times 1$ vector of latent common
  factors representing inherent pathway effects for sample $i$.
  \item $ \Lambda$ is a $p \times q$ factor loadings matrix in the CFA model,
  relating gene expression to pathway effects. $\Lambda_k$ is the $k$-th row
  of $\Lambda$.
  %\item $B$ is the $q \times q$ matrix of regression coefficients in the CAR
  %model.
  \item $\rho_{ji}$ is a random ``external'' perturbation effect for pathway
  $j$ in sample $i$.
  \item $\theta_{ji}$ is an indicator for pathway $j$ in sample $i$ being
  perturbed.
	\end{itemize}
	
We describe and motivate the various aspects of this model in more detail in
the following subsections.

\subsection{Modelling Biological Replicates}
It is common in biological studies to have replicates. In this case we allow
perturbations and latent factors to vary for each replicate, yielding the
model
\begin{align*}
y_{kir} \given \Lambda, \omega_{ir}, \psi_k &\ind
N(\Lambda_k \omega_{ir}, \psi_k) \tag{\ref{cfa_model}'} \\
\omega_{jir} \given \omega_{[-j]ir}, \rho_{jir}, \sigma^2 &\ind
N\Bigg( \sum_{j'=1, j' \neq j}^{q} B_{jj'} \omega_{j'ir}
+ \rho_{jir}, \sigma^2\Bigg) \tag{\ref{car_model}'} \\
\rho_{jir} \given \theta_{ji}, \tau^2 &\ind
N(0, \theta_{ji} \tau^2 + (1 - \theta_{ji}) v_0 \tau^2)
\tag{\ref{vs_model}'}
\end{align*}
for replicates $r = 1, \ldots, m_i$ in each study. Note that while $\rho$ and
$\omega$ vary, we assume the same perturbation targets and gene-pathway
loadings across replicates and so $\theta$ and $\Lambda$ do not depend on
index $r$. In the next sections we present each of the these levels in detail,
but we drop the replicate indices to simplify the notation.

\subsection{Confirmatory Factor Level}
\label{sec:cfa}
In the first component, we focus on the likelihood of individual gene
expression measurements.  It has been shown that the transcriptional activity
of a gene is, in part, regulated by  pathways positioned upstream of gene
regulation (e.g. \citep{vogelstein}). We describe this relationship between a
gene and its pathways using a \emph{confirmatory factor analysis} model, as
shown in Eq.~(\ref{cfa_model}).  Here, the gene expression measurements, $Y$,
are regressed on a set of biologically structured latent factors $\omega$
(biological pathways).  

Both the structure of the latent factors $\omega$ and the dependency between
individual genes $y$ and these latent factors are encoded, a priori, in the
$p \times q$ loading factor matrix $\Lambda$.  We integrate biological
information into the model through $\Lambda$, by limiting some of the elements
of $\Lambda$ to zero, where the non-zero elements in each \emph{column} of
$\Lambda$ correspond to the genes of a known canonical pathway.  That is, a
gene $k$ loads onto a pathway $j$, (that is, $\lambda_{kj} \neq 0$) if gene
$k$ (or its gene product) is a member of pathway $j$.  If a gene (or its gene
product) is not in a pathway, the corresponding loading factor is constrained
to $0$. Thus genes are regressed only on the pathways to which they belong.

We use the following priors for the loading factors $\Lambda$:
\[
\lambda_{kj} \iid \Bigg\{
\begin{array}{l} N(0, 0.1), \mbox{ if gene $k$ is in pathway $j$} \\
\delta_0(\cdot), \mbox{ otherwise} \end{array}
\]
where we set the variance of $\Lambda$ to be small to prevent the latent
factors from shrinking towards zero in practice, and keep the pathway noise
variance weakly informative.

We give the gene variances $\Psi$ inverse gamma priors,
\begin{equation*}
%\label{psi_prior}
\psi_k \iid \mbox{\sf IG}(\zeta, \zeta - 1), \quad k = 1, \ldots, p,
\end{equation*}
where the shape is chosen so that the prior mean is~$1$. We now control the
prior strength of the gene noise variance parameters $\Psi$
through the hyper-parameter $\zeta$ to avoid overfitting the model. This prior
regularization is important in cases where the data size is not large enough
to fit complex hierarchical models and can lead to instability, poor
convergence, multiple local modes, and empirically non-identified models when
the prior on each $\psi_k$ is too flexible. The extent to which an inverse
gamma prior is informative depends on the strength of the data and is
translated as a variance tradeoff between the prior and the data in the
conditional posterior distribution of $\psi_k$:
\begin{equation}
\label{eq:psi}
\psi_k \given \Omega, \Lambda, Y \sim
\mbox{\sf IG}\Bigg(
\zeta + \frac{n}{2},
\zeta - 1 + \frac{1}{2}\sum_{i=1}^{n}{(Y_{ki} -
\Lambda_{k,\cdot}\omega_{i})}^2
\Bigg).
\end{equation}

Since an inverse gamma distribution with shape $\alpha$ corresponds to a
$\chi^2$ distribution with $2\alpha$ degrees of freedom \citep{gelman_2006},
the prior provides $2\zeta$ pseudo-observations and the conditional
in~(\ref{eq:psi}) has $2\zeta + n$ observations. We thus constrain the prior
variances of $\Psi$ by setting $\kappa$ times data observations as prior
pseudo-observations, $\zeta = \kappa n / 2$, and so the conditional
in~(\ref{eq:psi}) has $n(\kappa + 1)$ degrees of freedom. In this manner, we
fix $\kappa < 1$ so that the prior does not overpower the likelihood.  In our
implementation, we fixed $\kappa$ to $0.5$ to have a reasonably informed
prior.

\subsection{Conditional Autoregressive Level}
The second component of the model is designed to characterize a perturbed
system of interacting pathways. The dependency between pathways is captured
by a \emph{conditional autoregressive model} \citep{cressie,oliveira} in
Eq.~(\ref{car_model}). The reasoning is that, under normal stationary
conditions, each pathway can be explained, on average, by other pathways
in a pathway-pathway network through the coefficients~$B$. However, when a
pathway is targeted by some external perturbation source (e.g.\ from disease,
or drug perturbation), then the expected mean of this pathway is shifted by an
additional term $\rho$ due to the perturbation. In the context of drug
perturbations, one can effectively imagine that after a drug hits a specific
pathway, this drug-induced effect propagates across the network, affecting
many pathways by mere association, and altering cellular phenotype. 

In the CAR level of the model, Eq.~(\ref{car_model}), the $q \times q$
design matrix $B$ represents a system of pathway-pathway interactions.
More specifically, let us initially set $B := \gamma W$, where $\gamma$ is
referred in the CAR literature as the \emph{spatial scaling} parameter, and
$W$ is a symmetric zero-diagonal matrix. We briefly describe the pathway
network $W$ in Section~\ref{pathway_network_description} and provide a
technical description in Appendix~\ref{A.network_construction}.

If we define $\Phi := {(I-\gamma W)}^{-1}$, where $I$ represents the identity
matrix of order $q$, the joint distribution of $\omega$ can be written as the
following \citep{cressie}:
\begin{equation}
\label{omega_prior}
\omega \given \rho, \Phi, \sigma^2 \sim N( \Phi \rho, \sigma^2 \Phi).
\end{equation}
Thus, if $\gamma = 0$ the CAR model is reduced to an exploratory factor
analysis model. However, to attain an identifiable model (please refer to
Section~\ref{model_ID} for details), we require that the factors
have equal variance $s^2$ and so $\Phi = s^2 R(\gamma)$, where $R(\gamma)$ is
the correlation induced by $G(\gamma) := {(I - \gamma W)}^{-1}$,
\begin{equation}
\label{Phi_equation}
R(\gamma) = \mbox{\sf Diag}{(G(\gamma))}^{-\frac{1}{2}}
G(\gamma)
\mbox{\sf Diag}{(G(\gamma))}^{-\frac{1}{2}}.
\end{equation}

In addition, to ensure that $\Phi$ is positive definite, we constrain $\gamma$
to the interval $(1/\iota_1, 1/\iota_q)$, where $\iota_1$ and $\iota_q$ are
the minimum and maximum eigenvalues of $W$, respectively. The prior on
$\gamma$ is uniform,
\[
\gamma \sim \mbox{\sf Unif}\Bigg(
\frac{1}{\iota_1} + \delta, \frac{1}{\iota_q} - \delta
\Bigg),
\]
where $\delta > 0$ ensures that the maximum correlation in $R(\gamma)$ is not
near $1$, which can lead to a non-identified model.  We set $\delta = 0.005$,
which, in our experience, bounds the maximum correlation at roughly $0.90$.
Finally, we set a weakly informative prior on $\sigma^2$,
$\sigma^2 \sim \mbox{\sf IG}(0.001, 0.001)$.

\subsection{Interaction Network in the CAR Model}
\label{pathway_network_description}

The pathway interaction network lies at the core of our model.   We
specifically use a network, where nodes are pathways, constructed in a manner
that implicitly links pathways by their common function in the cell.  To date
there are several ways of building a pathway interaction network.  The most
ad-hoc approach is to simply define a link between two pathways where there is
a non-empty intersection of gene members.  However, this rule overlooks
interactions that may occur \emph{between non-overlapping genes} of two
pathways.  An alternative approach is to use a protein-protein interaction
database to identify physical interactions \emph{between} members of two
pathways, and using an aggregate score to define the overall interaction link.
However, using PPI interactions is often very noisy given the high variability
in PPI data \citep{von_2002,reguly_2006,gandhi_2006} resulting from methods
that produce high-coverage of the proteome inherently having a high false
positive rate.

To better capture interactions between pathways, we used Gene Ontology's (GO)
biological processes \citep{go} to define functional links between pathways.
The advantage of GO over a PPI database is that a GO gene set is manually
curated, thus containing much fewer false positives.  Functional networks have
been used extensively in biology \citep{ideker_2002,ideker_2007,ideker_2008},
which are generally motivated by the principle that genes that work together
to accomplish a task in the cell are usually co-activated/inhibited or
co-\@(de)\@regulated.

To construct the pathway-pathway interaction network $W$, we regard the set of
pathways as a weighted network where the nodes represent canonical pathways
and the edge weights in $W$ reflect the degree of functional similarity
between two pathways. As an example, in Fig.~\ref{MoA_illustration} the nodes
would represent Pathways A, B, and C with a link connecting Pathways A to B
and A to C. We then create $W$ in a manner similar to an algorithm by
\citet{lpia} using KEGG regulatory and signaling pathways \citep{kegg2,kegg3},
and GO biological processes \citep{go} obtained from the MSigDB collection
\citep{gsea}. For a technical description of the network construction see
Section~\ref{A.network_construction}.

\subsection{Spike-and-Slab Prior}
Our goal of elucidating primary targets of a perturbation reduces to
identifying perturbation effects after accounting for all pathway-pathway
interactions (i.e.\ via network filtering) as described by a known biological
network $W$. Importantly, we assume that relatively few pathways are primary
targets of a perturbation.  We use a spike-and-slab prior on $\rho$
\citep{george} and identify perturbations through posterior-based variable
selection.  Thus our parameters of interest are variables $\theta_{ji}$ that
indicate whether $\rho_{ji}$ represents a non-zero perturbation for pathway
$j$ in sample $i$, as in Eq.~(\ref{vs_model}).

Because the scale of the perturbations is unknown, we choose a mixture of two
normals \citep{george}, one approximating the spike and the other the slab, in
such a way that is invariant to the scale of the perturbations.
That is, we define the spike variance to be some fraction $v_0$ of the slab
variance \citep{piette} and let the slab variance be random. We fixed
$v_0 =0.01$, meaning that a non-perturbed pathway has only one-hundredth of
the variance of a perturbed pathway, while keeping the prior on $\tau^2$
weakly informative, similarly to the prior on $\sigma^2$.

Lastly, the prior probability of a pathway being perturbed is captured by a
hyper-parameter $\alpha$. We expect $\alpha$ to be small to reflect the
targeted effect of drug perturbations to a few pathways because, for instance,
the FDA requires perturbation selectivity for drug approval. Thus, to induce
sparsity we set $\alpha = 0.1$ as an upper bound for the expected fraction of
perturbed pathways, yielding:
\begin{align*}
\theta_{ji} &\iid \mbox{\sf Bern}(0.1) \\
\tau^2 &\sim \mbox{\sf IG}(0.001, 0.001),
\end{align*}
for $j = 1, \ldots, q$ and $i = 1, \ldots, n$. While we expect \emph{a priori}
that at most $10\%$ of the pathways are selectively perturbed, we found in a
prior sensitivity analysis that small changes in $\alpha$ do not significantly
change our inference on perturbation detection.

\subsection{Model Identification}
\label{model_ID}

Model identification has always been a non-trivial task with factor models
(see Chapters~4, 7, 8 in \citep{bollen}). In CFA models, constraints are
usually applied to elements of both $\Phi$ and $\Lambda$. This can be done,
for example, by fixing the factor variances and the first $q$ rows of
$\Lambda$ to a particular structure \citep{west_2006}. These methods prevent
non-trivial rotational or scale transformations of $\Lambda$ and $\Phi$ by
forcing the transformation matrix U to be the identity matrix. However, since
in our model $\Phi := s^2 R$, where $R$ is a correlation matrix
(Eq.~(\ref{Phi_equation})), $\Phi$ is already rotationally unique.  

This model, however, is still not identified in a few additional ways.
Firstly, for the $i$-th sample we have $Y_i \given \Lambda, \omega_i, \Psi
\sim N(\Lambda \omega_i, \Psi)$ and so, if we marginalize $\omega_i$
we have the following conditional distribution on $Y_i$:
\begin{equation}
\label{eq:condYi}
Y_i \given \Lambda, \rho_i, \sigma^2, \gamma, \Psi \sim
  N(s^2 \Lambda R(\gamma) \rho_i,
    \Psi + \sigma^2 s^2 \Lambda R(\gamma) \Lambda^\top).
\end{equation}
Thus, if we re-scale $\tilde{\sigma}^2 = k_\sigma \sigma^2$,
$\tilde{s}^2 = k_s s^2$, $\tilde{\rho}_i = k_\rho \rho_i$, and
$\tilde{\Lambda} = k_\Lambda \Lambda$ such
that
\[
k_s k_\Lambda k_\rho = 1 \quad \mbox{and} \quad k_\sigma k_s k_\Lambda^2 = 1
\]
we obtain the same conditional distribution on $Y_i$ since its mean and
variance remain unaltered, respectively.
For example, re-scaling $\Lambda$, $\sigma^2$ and $\rho_i$ with any
scalar $a \ne 0$ such that
$\tilde{\Lambda} = a \Lambda$, $\tilde{\sigma}^2 = \sigma^2 / a^2$,
and $\tilde{\rho}_i = \rho_i / a$ does not affect the distribution of
$Y_i$.
To solve these forms of re-scaling, we fix $s^2 = 1$ as well as the prior
variance of $\Lambda$ to $0.1$, a small value that avoids shrinking the scale
on $\omega$ towards $0$ in practice.

Finally, there is an additional case that can lead to a non-identified model.
This last case arises from the latent factor correlation matrix $R$.
In the simplest case, suppose $\gamma \neq 0$ and $W$ is a connected network
(i.e. $R$ can only be arranged as a single block matrix). Under these
conditions, there exist exactly two solutions where
\[
\tilde{\Lambda} = -\Lambda, \quad \tilde{\omega} = -\omega,
\quad \tilde{\rho} = -\rho.
\]

This is simply a sign flip across all pathways, which would not affect the
underlying factor covariance structure $\Phi$.  In fact, if $R$ is rearranged
into a block diagonal matrix, then each set of pathways corresponding to a
block in $R$ can be flipped in this manner without affecting $\Theta$ and
yielding the same joint posterior density. In other words, if $R$ is
rearranged into a block diagonal matrix with $b$ blocks then there exist
exactly $2^b$ solutions. Therefore, perturbations are identified up to their
signs; that is, we can infer if two pathways are perturbed in the same
direction, but if the perturbation is positive or negative is arbitrary.
Moreover, since our goal is to infer perturbations, we are more interested in
the magnitude of a perturbation---whether it is significantly close to zero or
not---rather than its sign.

\section{Posterior Inference}
\label{sec:sampling}

Our primary goal is the posterior inference on $\Theta = \{\theta_{ji}\}$,
which is used to identify perturbation targets.   We obtained posterior
estimates to our model parameters via a partially collapsed hybrid Gibbs
sampler \citep{dyk} with an adaptive Metropolis step \citep{gelman}.
We call the sampler ``partially collapsed'' because we sample some of the
parameters (namely, $\rho$ and $\theta$) from a \emph{marginalized}
conditional posterior. We found that using this partially collapsed sampler
facilitates the sampler to move into regions of high posterior mass
concentration faster.

First, the steps for an ordinary, non-collapsed Gibbs sampler (dropping
iteration indices and irrelevant conditional parameters) are:
\begin{align*}
&[\Lambda \given \omega, \psi, Y], &
%[&\rho \given \Lambda, \theta, \psi, Y], & % collapsed
&{[\rho \given \theta, \Phi, \omega, \sigma^2, \tau^2]}^*, &
%[&\theta \given \omega, \Phi, \tau^2], \\ % collapsed
&{[\theta \given \rho, \tau^2]}^*, \\
&[\gamma \given \omega, \rho, \Phi, \sigma^2], &
&[\omega \given \Lambda, \rho, \psi, \Phi, \sigma^2, Y], &
&[\psi \given \Lambda, \omega, Y], \\
&[\sigma^2 \given \omega, \rho, \Phi], &
&[\tau^2 \given \rho, \theta].
\end{align*}
Note that, even though $\Phi = s^2 R(\gamma)$ is a matrix that depends on
$\gamma$, we use them interchangeably to make the conditional distributions
clearer.
Now, to improve the rate of convergence, we marginalize the two steps starred
above: (i) instead of sampling from
$[\rho \given \theta, \Phi, \omega, \sigma^2, \tau^2]$, we integrate out
$\omega$ from
$[\rho, \omega \given \Lambda, \theta, \Phi, \psi, \sigma^2, \tau^2, Y]$; and
(ii) instead of sampling from $[\theta \given \rho, \tau^2]$, we marginalize
$\rho$ from $[\theta, \rho \given \omega, \Phi, \tau^2, \sigma^2]$. The
collapsed sampler has then updated steps
\[
[\rho \given \Lambda, \theta, \Phi, \psi, \sigma^2, \tau^2, Y]
\quad \mbox{and} \quad
[\theta \given \omega, \Phi, \tau^2, \sigma^2].
\]
We note that, as \citet{dyk} pointed out, marginalizing does not alter the
stationary distribution of the full posterior nor the compatibility of the
conditional distributions.
In the next sections we provide details on each of these steps.

\subsection{Initializing MCMC chains}
We implemented a tempering algorithm at the beginning of our MCMC chains to
find reasonable starting points such that the chains were less likely to get
stuck at a local mode.  These local modes are caused, in part, by each
likelihood in the CFA level, Eq.~(\ref{cfa_model}), being close to
non-identifiable up to the mean $\Lambda \omega$.
To do this, we run our sampler as usual, but for $\Psi$, $\Lambda$, and
$\Omega$, we temper the likelihood distributions
$Y \given \Psi, \Lambda, \Omega$ when obtaining their conditional posteriors.
Similarly, we do the same for the likelihood of $Y \given \rho$ when sampling
$\rho$.  At hot temperatures, the likelihood of $Y$ would be flatter, allowing
the chains to move more freely around the space. We slowly cool the
temperature to $T=1$ to obtain the targeted posterior and thus beginning the
true sampler.

\subsection{Sampling $\Lambda$}

Recall that we have constrained some elements in $\Lambda$ to $0$ such that
genes are regressed only on pathways to which they belong.  That is, if gene
$k$ is not in pathway $j$, then we constrain the element $\lambda_{kj}$ to
$0$.  Consider the $k$-th row of $\Lambda$, which we denote as $\Lambda_k$, and
suppose some of the elements in $\Lambda_k$ are constrained to zero.  Let
$c_k$ be the corresponding $1 \times q$ row vector such that
\[
c_{kj} := I(\mbox{gene $k \not\in$ pathway $j$}),
\]
that is, $c_{kj} = 0$ if $\lambda_{kj}$ is a structural zero and so
$\lambda_{kj} \iid N(0, (1-c_{kj})0.1)$.
If $r_k = \sum_j c_{kj}$ is the number of pathways containing
gene $k$ and $\Lambda_k^{*}$ is the $1 \times r_{k}$ row vector that
contains the unknown parameters in $\Lambda_k$ then the prior on $\Lambda_k^*$
is $N(0, H_k)$ where $H_k = 0.1 I$, $I$ the identity matrix of order $r_k$.

Similarly, let $\Omega_k^*$ be the $r_k \times n$ sub-matrix of $\Omega$ such
that for $j = 1, \ldots, r_k$, all the rows corresponding to $c_{kj} = 0$ are
deleted and also let $Y_{k,\cdot}$ be the $1 \times n$ vector of observations
for gene $k$. Then, from a well known result of \citet{lindley}, the posterior
conditional distribution for $\Lambda_k^*$, $k = 1, \ldots, p$, is
\[
\Lambda_k^* \given Y_{k}, \Omega_k^*, \psi_k \sim N(A_k^{-1} a_k, A_k^{-1})
\]
where $A_k = \psi_k^{-1} \Omega_k^* {\Omega^{*}_k}^\top +  H_k^{-1}$ 
and $a_k = \psi_k^{-1} \Omega_k^* Y_{k,\cdot}^\top$.

\subsection{Sampling $\rho$}
\label{rho_posterior}

We define $\Sigma(\theta_i)$ as a diagonal matrix whose $j$-th diagonal
element is
${\Sigma({\theta_i})}_{jj} = \tau^2[\theta_{ji} + v_0 (1-\theta_{ji})]$.
Then, again exploiting the result of \citet{lindley} on the marginalized
conditional distribution of $Y_i$ after integrating out $\omega_i$
in~(\ref{eq:condYi}), we sample $\rho_{i}$ from the
following conditional posterior distribution:
\[
\rho_i \given \theta_i, \Lambda, \Psi, Y_i \sim
N( A_i^{-1}a_i, A_i^{-1}),
\]
where
\begin{align*}
A_i &= \Phi^\top \Lambda^\top
{(\sigma^2 \Lambda \Phi \Lambda^\top + \Psi)}^{-1}
\Lambda \Phi + {\Sigma(\theta_i)}^{-1}, \\
a_i &= \Phi^\top \Lambda^\top
{(\sigma^2 \Lambda \Phi \Lambda^\top + \Psi)}^{-1} Y_{i}.
\end{align*}

\subsection{Sampling $\theta$}

As in the previous section, we compute the marginalized conditional posterior
of $\theta$ after integrating out $\rho$. For computational efficiency, we
store the marginal variance $V(\theta_i^{(t)})$ of $\omega_i$ conditional on
$\theta_i$ at each iteration $t$ of our sampler, where index $i$ runs over
samples. That is,
\[
V(\theta_i) = \sigma^2 \Phi + \Phi \Sigma(\theta_i) \Phi^\top.
\]
To simplify the notation, we drop the sample index $i$ on all parameters
including $\theta$.

At this step we sample iteratively from
$\theta_j \given \theta_{[-j]}, \omega, \Phi, \tau^2$ for $j = 1, \ldots, q$.
There are then two cases: when $\theta_j^{(t)} = 0$ and when
$\theta_j^{(t)} = 1$, for which we define
$V_0 = V(\theta_j^{(t)}=0, \theta_{[-j]}^{(t)})$ and
$V_1 = V(\theta_j^{(t)}=1, \theta_{[-j]}^{(t)})$.
Let $\phi_j$ denote the $j^{th}$ column of $\Phi$.  If $\theta_j^{(t)} = 0$,
then we define
$\delta_{j0} = 1 + (\tau^2 -v_0\tau^2) \phi_j^\top V_0^{-1} \phi_j$ and
$\Delta_{j0} = (\tau^2 - v_0\tau^2) / \delta_{j0}
(V_0^{-1}\phi_j){(V_0^{-1}\phi_j)}^\top$ to obtain:
%
%\begin{multline*}
\[
\logit\, \Pr\Big(\theta_j^{(t+1)} = 1 \given \theta_{[-j]}, \omega, \Phi\Big)
= %\\
-\frac{1}{2} \log \delta_{j0}
+ \frac{1}{2} \omega^\top \Delta_{j0} \omega
+ \logit(\alpha).
\]
%\end{multline*}

If $\theta_j^{(t)} = 1$ we define, similarly,
$\delta_{j1} = 1 - (\tau^2 - v_0\tau^2) \phi_j^\top V_1^{-1} \phi_j$
and $\Delta_{j1} = (\tau^2 - v_0\tau^2)/\delta_{j1} (V_1^{-1} \phi_j)
{(V_1^{-1} \phi_j)}^\top$ to obtain:
%
%\begin{multline*}
\[
\logit\, \Pr\Big(\theta_j^{(t+1)} = 1 \given \theta_{[-j]}, \omega, \Phi\Big)
= %\\
\frac{1}{2}\log\delta_{j1}
+ \frac{1}{2}\omega^\top \Delta_{j1} \omega
+ \logit(\alpha).
\]
%\end{multline*}
%
For the full derivation of these posterior probabilities, please refer to
Appendix~\ref{A.theta}.

\subsection{Sampling $\gamma$}

As with most spatial models, the scaling parameter $\gamma$ lacks conjugacy.
We used a random walk adaptive Metropolis algorithm such that the proposal
density is normal and centered at the current sample, $\gamma^{(t)}$, with
variance $\xi^2$. This variance is tuned to adjust for the acceptance
rate and fixed post burn-in.

We propose $\gamma^* \sim N(\gamma^{(t)}, \xi^2)$.  Let $\Phi^* = s^2
R(\gamma^*)$ with the correlation computed as in Eq.~(\ref{Phi_equation}).
Then, if
\[
l(\Phi) := -\frac{n}{2}\log|\Phi|
-\frac{1}{2} \sum_{i=1}^n {(\omega_i - \Phi \rho_i)}^\top
{(\sigma^2 \Phi)}^{-1} (\omega_i - \Phi\rho_i),
\]
the acceptance ratio of the Metropolis step is just
$r(\gamma^*, \gamma^{(t)}) = \exp\{l(\Phi^*) - l(\Phi^{(t)})\}$
since we have a flat prior on $\gamma$.

\subsection{Sampling $\omega$, $\psi$, $\sigma^2$, and $\tau^2$}
We sampled $\omega$ according to the linear conditional posterior:
\[
\omega_i \given Y_i, \Lambda, \Psi, \rho_i, \gamma \ind
N(A_i^{-1} a_i, A_i^{-1})
\]
where
\begin{align*}
A_i &=
\Lambda^\top \Psi^{-1} \Lambda + \sigma^{-2}\Phi^{-1} \\
a_i &=
\Lambda^\top \Psi^{-1} Y_i + \sigma^{-2}\rho_i.
\end{align*}

The conditional distributions of variance parameters $\psi$ and $\sigma^2$
follow from conjugacy:
\begin{align*}
\psi_k \given \omega, \Lambda, Y &\sim
\mbox{\sf IG}\Bigg(\zeta + \frac{n}{2},
\zeta - 1 + \frac{1}{2}\sum_{i=1}^{n}{(Y_{ki} - \Lambda_{k,}\omega_{i})}^2
\Bigg) \\
\sigma^2 \given \omega, \Phi, \rho, &\sim
\mbox{\sf IG}\Bigg(0.001 + \frac{qn}{2}, \\
& \quad
0.001 + \frac{1}{2}\sum_{i=1}^{n} {(\omega_{i} - \Phi \rho_{i})}^\top
\Phi^{-1} (\omega_{i} - \Phi \rho_{i}) \Bigg).
\end{align*}
where $\zeta = n / 4$ according to the discussion in Section~\ref{sec:cfa}.

The conditional distribution of $\tau^2$ is also conjugate; if $v_{ij} = 1$
when $\theta_{ij} = 1$ and $v_{ij} = v_0$ when $\theta_{ij} = 0$, then
\[
\tau^2 \given \rho, \theta \sim \mbox{\sf IG}\Bigg(
0.001 + \frac{qn}{2},
0.001 + \sum_{i=1}^{q} \sum_{j=1}^{n} \frac{\rho_{ij}^2}{2v_{ij}} \Bigg).
\]

\subsection{Posterior Inference}
We infer perturbations based on the centroid estimator of $\theta$ for some
threshold $t$ \citep{carvalho}:
\begin{equation}
\label{thetahat}
\hat{\theta}_{ji}(t) := I\Big(\Pr(\theta_{ji} \given Y) > t\Big).
\end{equation}
We choose the threshold $t$ by controlling a Bayesian false discovery rate. We
define the Bayesian false discovery rate (BFDR) of an estimator as
%
%\begin{multline}
\begin{equation}
\label{bfdr}
\mbox{BFDR}(\hat{\theta}(t)) :=
\Exp_{\theta \given Y} \Bigg[
\frac{\sum_{i=1}^n \sum_{j=1}^q \hat{\theta}_{ji} (1 - \theta_{ji})}
{\sum_{i=1}^n \sum_{j=1}^q \hat{\theta}_{ji}} \Bigg] %\\
= \frac{\sum_{i=1}^n \sum_{j=1}^q \hat{\theta}_{ji} (1 - \Pr(\theta_{ji} = 1
\given Y))}
{\sum_{i=1}^n \sum_{j=1}^q \hat{\theta}_{ji}}.
\end{equation}
%\end{multline}

\section{Simulation Study}
\label{sec:simulation}

We conducted a simulation study to (i) assess the impact of including
biological network information into a factor model, and (ii) test the
robustness of our model to inaccuracies in the biological databases used to
construct pathways and the pathway-pathway interaction network.  

\subsection{Assessing the impact of incorporating biological network
information}
\label{sim_1}

We tested the CFA-CAR model under various signal-to-noise (SNR) ratios and
compared its performance with an exploratory factor analysis (EFA) model.
This comparison model can be obtained from our original CFA-CAR model, by
letting the pathway-pathway network be empty.  That is, an EFA model is a
factor model such that the factor covariance matrix, $\Phi$, is
diagonal, which is equivalent to setting $\gamma = 0$ in the CFA-CAR model. In
effect, this constraint removes the network filtering effect of the CAR model.
The EFA model can thus be written as (Eq.~\ref{EFA}):

\begin{align}
\label{EFA}
y_{ki} \given \Lambda, \omega_i, \psi_k &\ind
N(\Lambda_k \omega_{i}, \psi_k)  \\
\omega_{ji} \given \omega_{[-j]i}, \rho_{ji}, \sigma^2 &\ind
N( \rho_{ji}, \sigma^2)  \\
\rho_{ji} \given \theta_{ji}, \tau^2 &\ind
N(0, \theta_{ji} \tau^2 + (1 - \theta_{ji}) v_0 \tau^2)
\end{align} 

We simulated toy-scale networks with graph densities similar to the original
KEGG pathway network.  For each network, we randomly selected $q = 10$
pathways from our set of KEGG pathways.  In this paper, we describe the
results obtained from one of these networks.   We obtained the pathway-pathway
weighted networks by taking the subnetwork from our original pathway-pathway
network induced by the ten randomly selected pathways.  The number of distinct
genes in the union of the selected pathways was 878.  We used the real gene
to pathway membership to create the mask of $\Lambda$.  We sampled the
non-zero loading factors in $\Lambda$ from a Gaussian with zero mean and
variance $0.1$.  The spatial scaling parameter $\gamma$ was relatively high to
emphasize non-trivial links in the network $W$.

We define the SNR in each simulated data set using the
SNR of a perturbed pathway (which is the same for any perturbed case).  If pathway $k$ was perturbed with effect $\rho$, we define
SNR$ =  \rho/\sigma$, and if we further fix the pathway noise variances ($\sigma$) to $1$, then the SNR$ =  \rho$. In each case sample, we simulated an experiment where only a single pathway was perturbed, with 5 ``biological'' replicates per experiment (perturbation). Therefore, if we perturbed pathway $k$ in sample
$i \in \{1,2,3,4,5\}$ in experiment $j$, we set $\rho_{ki}^{j} = m$, where $m$
is the signal to noise ratio in the data set, and $\rho_{k'i}^{j} = 0$ for $k'
\neq k$.  Across all
data sets, we fixed the gene variances ($\Psi$) to $1$.

For this network, we created six data sets of different signal to noise ratio in ten simulation replicates.  In each data set we fixed SNR to the following values: $0.50$, $1.50$, $2.50$, $3.50$, $4.50$, and $5.50$. These values were chosen based on the distribution of the SNRs in the NCI-DREAM drug sensitivity data sets (see Section~\ref{dream7} and Fig.~\ref{data_snr}).  For each of these data sets, we sampled 50 cases and 50 controls.

%For this network, we created six data sets of different signal to noise ratios in ten simulation replicates.  In each data set we fixed SNR to the following values: $0.50$, $1.50$, $2.50$, $3.50$, $4.50$, and $5.50$ by fixing $\sigma^2 = 1$ and varying the mean perturbation effect $\rho$. These values were chosen based on the distribution of the SNRs in the NCI-DREAM drug sensitivity data sets (see Section~\ref{dream7} and Fig.~\ref{data_snr}). For each of these data sets, we sampled 50 cases and 50 controls.     We calculated the SNR ratio in each data set using the SNR of a perturbed pathway (which is the same for any perturbed case).  That is, if pathway $k$ was perturbed with magnitude $m$, we computed SNR$ = s m/\sigma = m$. In each case sample, we simulated an experiment where only a single pathway was perturbed, with 5 ``biological'' replicates per experiment (perturbation).      fix the gene variances ($\Psi$) and

\begin{figure}[htbp]
\begin{center}
\includegraphics[width=\textwidth]{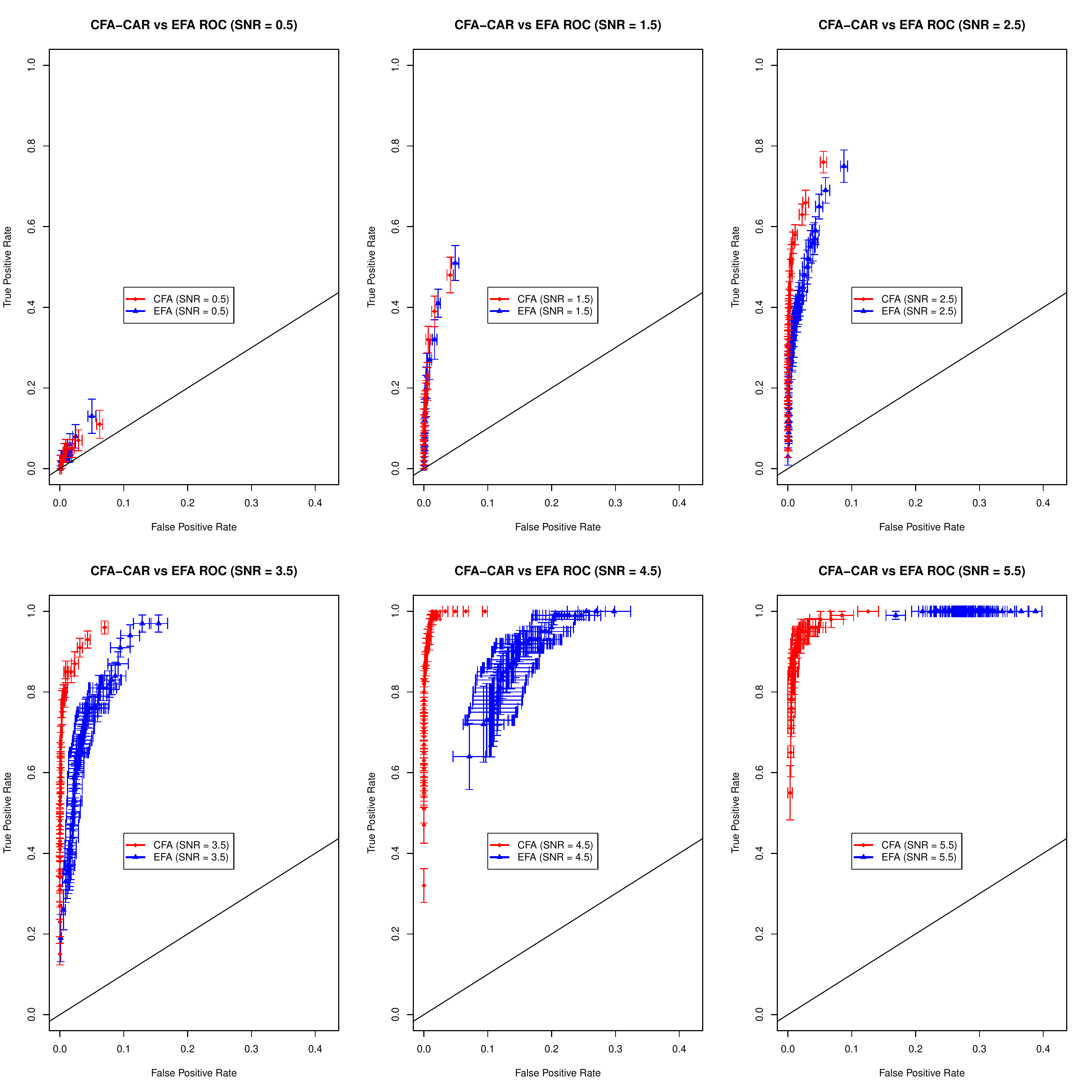}
\caption{Mean ROC curves with standard error bars
representing the standard errors across ten data replicates of the
CFA-CAR and EFA models. Each sub-figure corresponds to a different signal to
noise ratio in the data.  The solid line represents an expected ROC curve
under random guessing. Note that we truncated the x-axis at $0.42$ because
there were no additional plot points until $(x=1,y=1)$.}
\label{ROC_curves_part1}
\end{center}
\end{figure}

\begin{figure}[htbp]
\begin{center}
\includegraphics[width=0.5\textwidth]{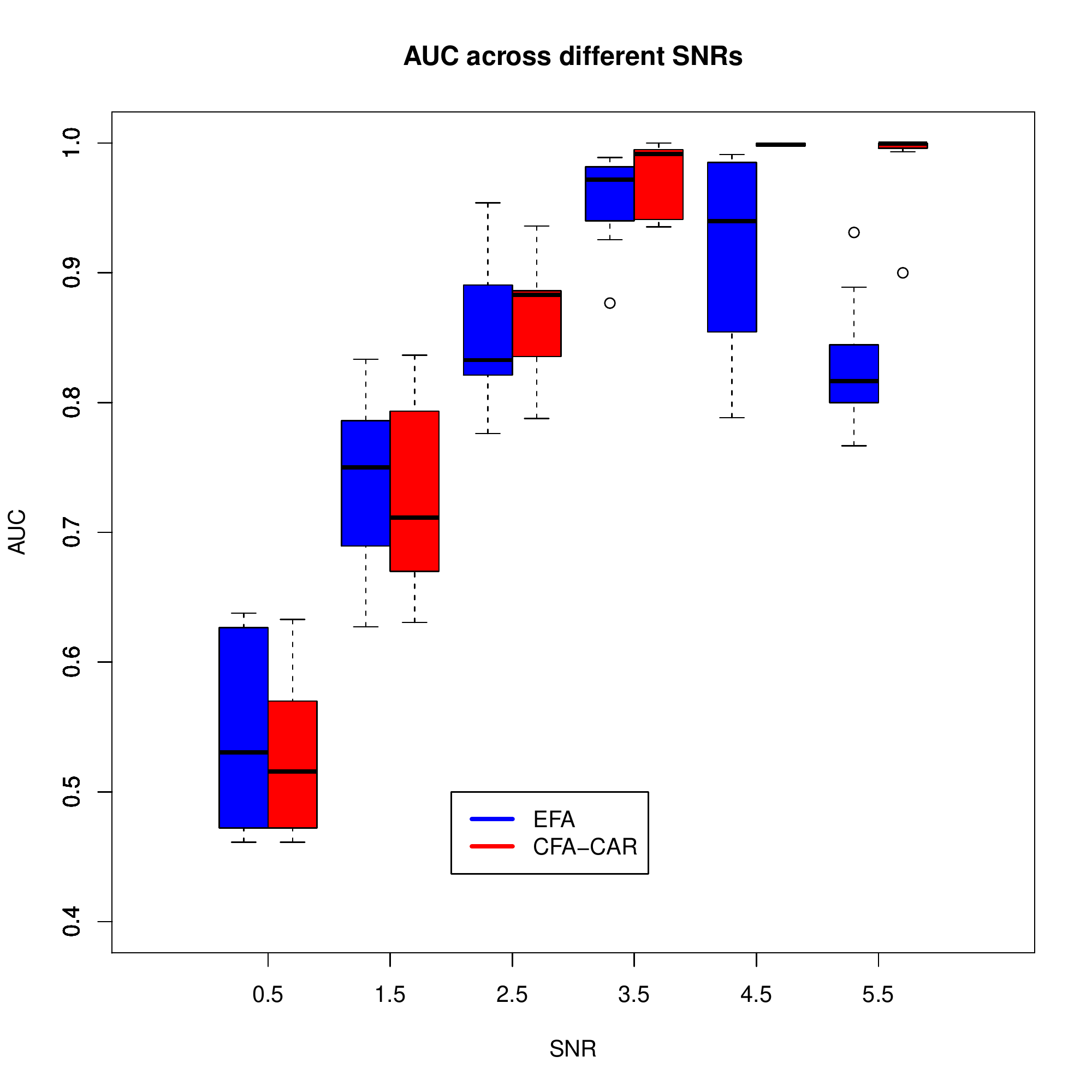}
\caption{Boxplot of the AUCs of the ROC plots in Fig.~\ref{ROC_curves_part1}.}
\label{ROC_AUC_part1}
\end{center}
\end{figure}

When running either models, we conditioned the perturbations of all
replicates, $\rho_{i}^{j}$, where $j$ indexes the experiment and $i$ indexes
the experimental replicate, on the same indicator vector $\theta_j$.  For the
controls, we fixed the corresponding indicators to zero.  We ran both our CFA
model and the EFA model on each data set.  Fig.~\ref{ROC_curves_part1} is a
series of ROC curves across data sets of different signal to noise ratios.
Both models performed similarly under low SNRs.  However, the curves begin to
separate as the signal in the data sets increased.  This is further emphasized
by the AUC boxplots in Fig.~\ref{ROC_AUC_part1}.    The CFA-CAR model
consistently exhibits a low false positive rate across varying signal to noise
ratios.  Unlike the CFA model, the false positive rate of the EFA model
increases as the signal to noise ratios increase.  When SNR $=4.5$ the AUC of
the EFA model begins to decrease dramatically.  This was expected because
given a strong enough perturbation, the EFA model confuses a perturbation with
the residual downstream effects of a perturbation.  That is, pathways
associated with the perturbed pathway are wrongly identified by the EFA model.   

\subsection{Assessing robustness to inaccuracies in the pathway database}

We performed sensitivity analysis on the CFA-CAR model to test how possible
inaccuracies in KEGG can affect the performance of CFA-CAR relative to an EFA
model.  With misannotated genes, the inaccuracy encoded in the mask of
$\Lambda$ would effect both the CFA-CAR and EFA models.  However, inaccuracy
in the pathway network would effect only the CFA-CAR model.  We provide an
example with the same network described in Section~\ref{sim_1} using a
simulated dataset with a signal to noise ratio of $3.5$.  

We tested six levels of gene set perturbations in the set of pathways, where
for each level, we randomly perturbed a percentage $x$ of genes.  This would
ultimately change the loading factor matrix $\Lambda$ as well as the
pathway-pathway network $W$.  For each level, we randomly chose $x \%$  of
genes represented in KEGG and reassigned them to new pathways.  

We perturbed $x= 1\%$, $2\%$, $4\%$, $8\%$, $16\%$, and $32\%$ of all genes in
our set of ten pathways.  For each of these perturbations, we ran ten
replicates where for each replicate we simulated a new ``pathway'' database
and ran CFA-CAR on the original data set, using the newly constructed
$\Lambda$ and $W$.

\begin{figure}[htbp]
\begin{center}
\includegraphics[width=0.75\textwidth]{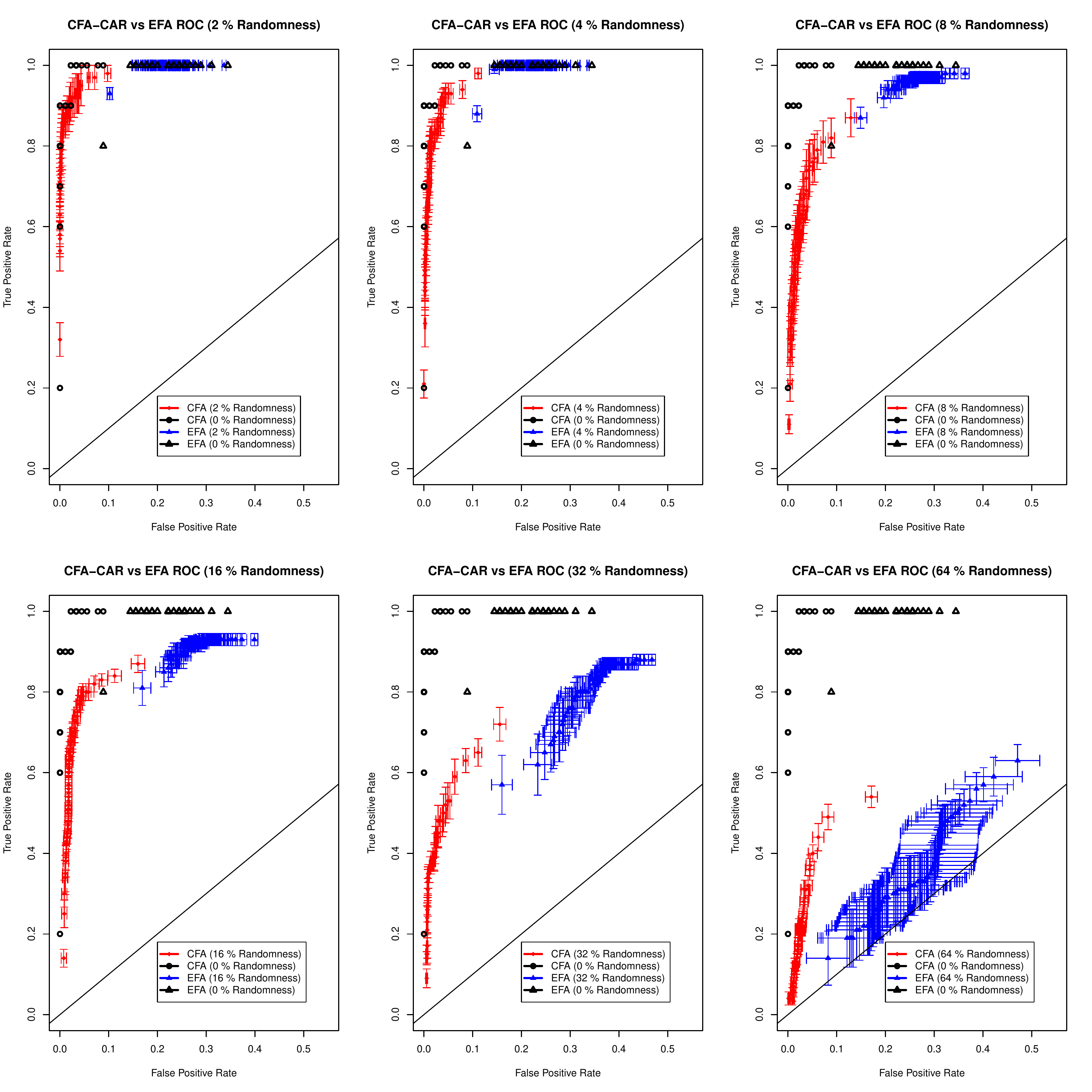}
\caption{Mean ROC curves with standard error bars for
the CFA-CAR model and the EFA model.  The standard error bars on the CFA-CAR
curve represent standard errors across different replicates of $W$ and
$\Lambda$.  The standard error bars on the EFA model represent standard errors
across different replicates of $\Lambda$.  Each sub-figure corresponds to a
different percentage of randomly reassigned genes.  The solid line represents
an expected ROC curve under random guessing.}
\label{ROC_curves_part2}
\end{center}
\end{figure}

\begin{figure}[htbp]
\begin{center}
\includegraphics[width=0.5\textwidth]{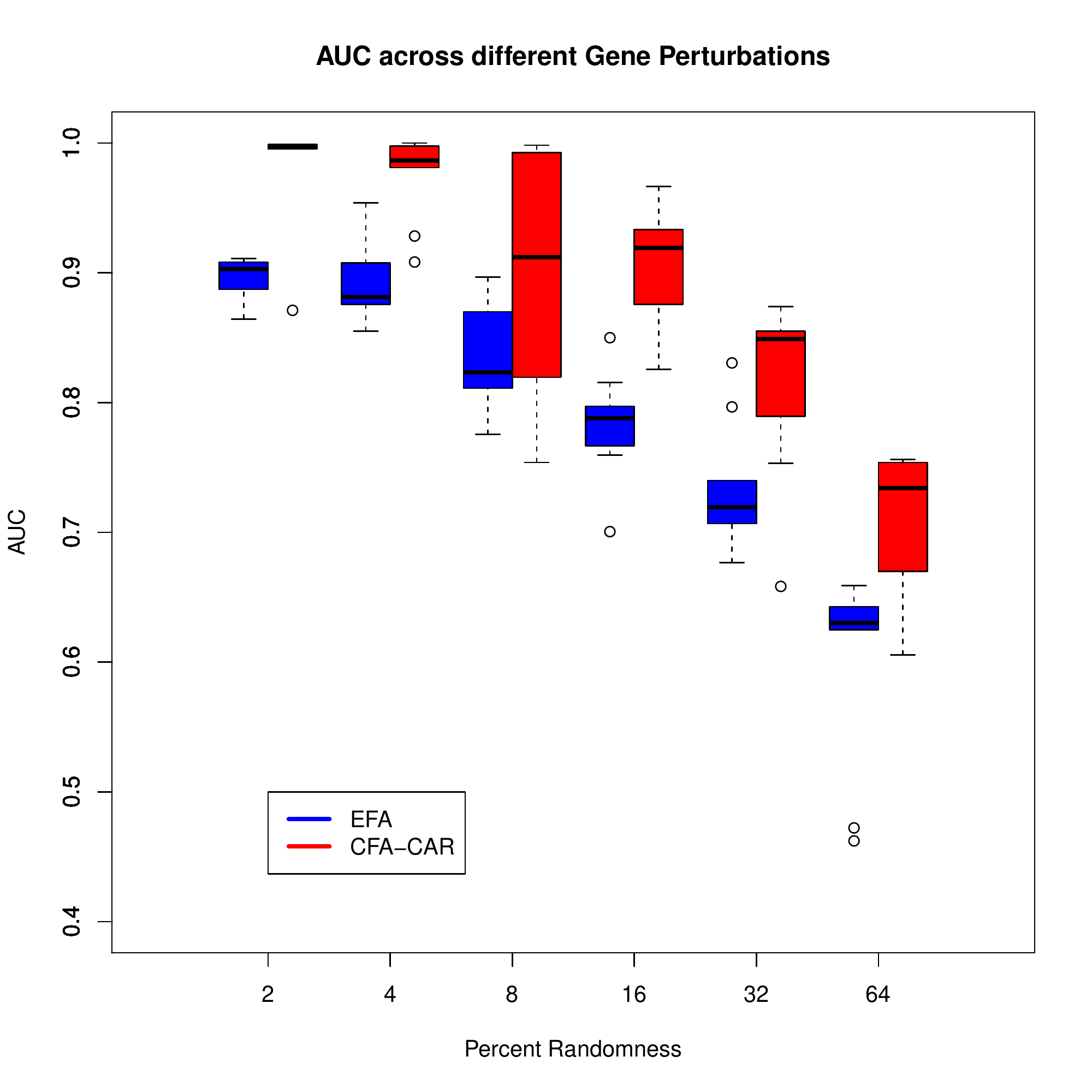}
\caption{Boxplot of the AUCs of the ROC plots in Fig.~\ref{ROC_curves_part2}.}
\label{ROC_AUC_part2}
\end{center}
\end{figure}

Fig.~\ref{ROC_curves_part2} is a series of ROC curves and
Fig.~\ref{ROC_AUC_part2} are AUC boxplots corresponding to the curves in
Fig.~\ref{ROC_curves_part2} for both the CFA-CAR and EFA model across various
levels of gene to pathway reassignments.  We begin to see a small departure
from the original ROC curve where the network and loading factor matrices were
not randomized when $8\%$ of the genes were randomly reassigned to different
pathways.  At this level, the specificity of the CFA-CAR model begins to
decrease.  However, even at $32\%$ randomness, the specificity and sensitivity
of the model is still significantly higher than what you expect to see by
random chance alone.  We conclude that the CFA-CAR model is rather robust to
possible inaccuracies in the biological databases, performing as well or
better than an EFA version of this model.

\section{Case Study: Drug Target Prediction in DREAM 7 Drug Sensitivity Challenge Data}
\label{dream7}

We applied the CFA-CAR model on the NCI-DREAM drug sensitivity challenge
data set.  As a comparison, we also implemented an EFA model (see Section
\ref{sec:simulation} for a description of the EFA model) and gene set
enrichment analysis (GSEA) \citep{gsea}.  The core algorithm of GSEA is a
threshold-free variant of a hypergeometric test; however, unlike the CFA-CAR
model, network interactions are not used.  

In the NCI-DREAM drug sensitivity data set, 14 compounds
(see Table~\ref{drug_list}) were tested at various concentrations and exposure
times on the LY3 (Lymphoma) cancer cell line.   We focused our attention on
the drug data sets with the strongest concentration (IC20) at either 12 or
24 hour exposures.  These data sets were compared to mock control data sets where
no drugs were exposed to the cells for 12 or 24 hour durations.  There were
eight mock replicates and each distinct experiment included three replicates,
yielding a total of 50 samples per data set (Data Set 1: IC20 at 24 hours
vs Mock at 24 hours and Data Set 2: IC20 at 12 hours vs Mock at 12 hours).

\begin{table}[htbp]
\begin{center}

\caption{Compounds tested on the LY3 cancer cell line in the NCI-DREAM data
challenge, the abbreviations used in the figures and description of drug and known mechanisms of action.}
\begin{tabular}{p{0.2\textwidth} p{0.1\textwidth} p{0.7\textwidth} }
\toprule
DRUG & Abbrev. & Description \\ \midrule
Aclacinomycin A & ACLA-A & Inhibits the degradation of ubiquinated proteins, affecting the proteasome \citep{acla} \\
Blebbistatin & BLEBB & Inhibits the myosin II protein affecting cellular motility pathways \citep{allingham} \\
Camptothecin & CPT & DNA-damaging agent targeting DNA topoisomerase I, affecting cell-cycle and p53-signaling \citep{gupta,jaks,wang} \\
Cycloheximide & CHX & Inhibits protein biosynthesis pathways \citep{chx} \\
Doxorubicin Hydrochloride & DOX & DNA-damaging agent targeting DNA topoisomerase II \citep{pommier}, and causes cell cycle arrest \citep{ling}.  Implicated in a P53-dependent mechanism of action \citep{ling,zhou,kurz}. \\
Etoposide & ETP & DNA-damaging agent causing DNA strands to break causing apoptosis and errors in DNA synthesis \citep{pommier}.  Involved in P53-dependent mechanisms of action \citep{karpinich,grandela} \\
Geldanamycin & GA  & Disrupts regulatory signaling mechanisms via HSP90 inhibition \citep{grenert,neckers}\\
H-7, Dihydrochloride & DHCL & Protein kinase C inhibitor \citep{dhcl} \\
Methotrexate & MTX & Inhibits DNA-synthesis via inhibition of purine and thymine synthesis \citep{goodsell}  \\
Mitomycin C & MTC\@. & DNA-damaging agent that is a potent DNA crosslinker \citep{thomasz} \\
Monastrol & MNS & Inhibits ATPase activity of the kinesin \citep{monastrol} \\
Rapamycin & RPM & Targets the mTOR protein \citep{alqurashi} affecting mTOR signaling pathway \\
Trichostatin A & TSA\@. &  Inhibits the class of histone deacetylase (HDAC) families of enzymes, inhibiting DNA transcription pathways \citep{trichostatin} \\
Vincristine & VCR & Causes cell-cycle arrest via inhibiting the assembly of microtubule structures during mitosis \citep{vincristine} \\

\bottomrule
\end{tabular}
\label{drug_list}
\end{center}
\end{table}

\begin{figure}[htbp]
\begin{center}
\includegraphics[width=\textwidth]{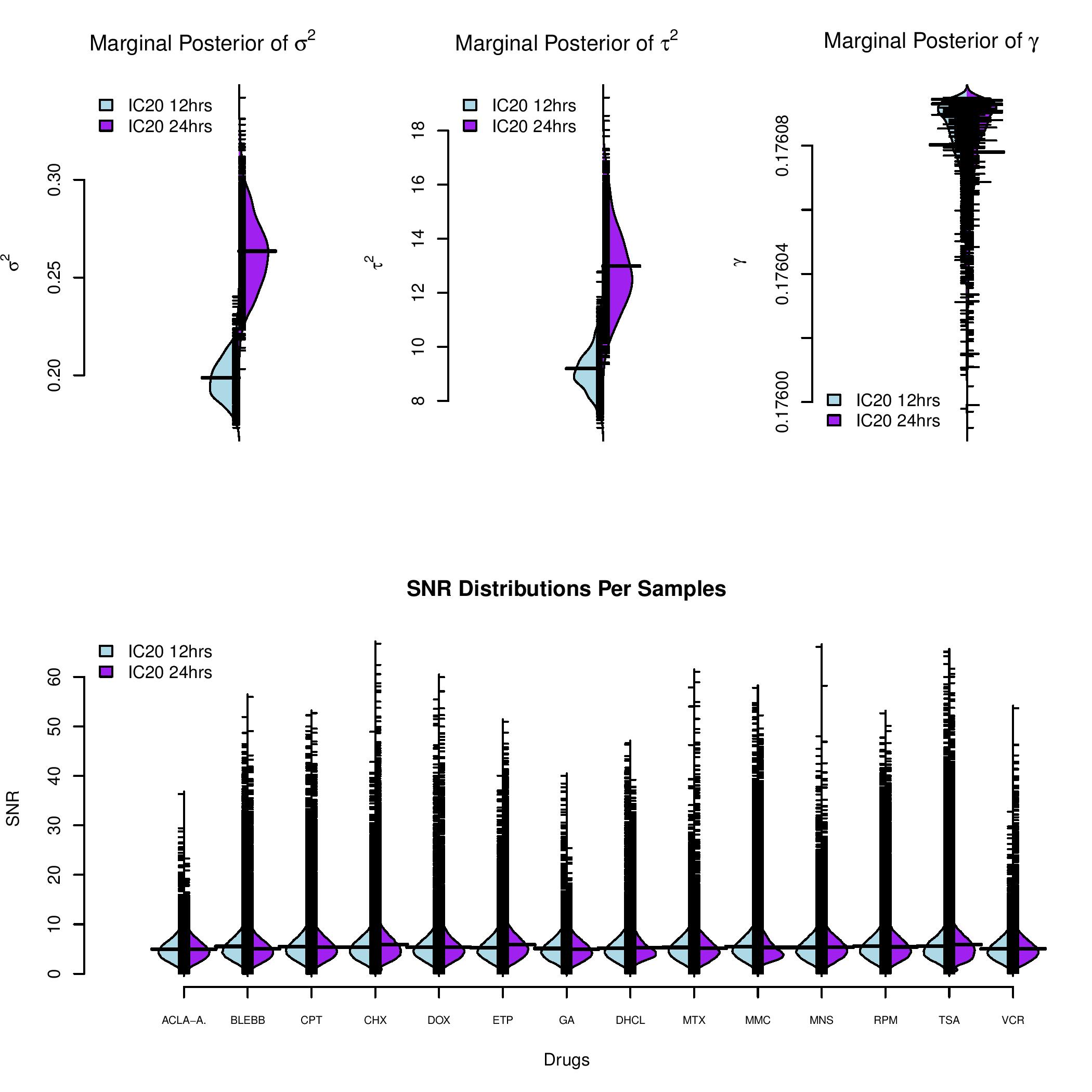}
\caption{Beanplots with bean averages of the marginal posterior distributions
of $\gamma$, $\sigma^2$, $\tau^2$, and the SNR across each drug for the IC20
at 12 hours and IC20 at 24 hours data sets.}
\label{data_snr}
\end{center}
\end{figure}

When running CFA-CAR and EFA models, we conditioned the perturbations
of all replicates, $\rho_1^{j}, \rho_{2}^{j}, \rho_{3}^{j}$, where $j$ indexes
the experiment, on the same indicator vector $\theta_j$.  In the control
cases, we fixed the corresponding indicators $\theta$ to $0$.  For each
dataset, we ran two chains of $4000$ iterations each, with a burn-in of
$2000$.  We assessed convergence using the Gelman-Rubin statistic on the
absolute values of all the continuous parameters.  Fig.~\ref{data_snr}
summarizes the posterior distributions of the model parameters $\gamma$,
$\sigma^2$, $\tau^2$, and the signal to noise ratios (SNRs) across the
different drugs for each data set (IC20 at 12 hours and IC20 at 24 hours).
We define the SNR of a pathway $j$ in a sample $i$ as the mean ratio, across
replicates, of the absolute perturbation $\rho$ to either $\sigma$, if the
pathway is perturbed, or $\sigma\sqrt{v_0}$ otherwise; that
is
$\mbox{SNR}_{ji} = \sum_{r=1}^{3} (\theta_{ji} |\rho_{jir}| / \sigma
+ (1 -\theta_{ji}) |\rho_{jir}| / (\sigma \sqrt{v_0})) / 3$,
%$\mbox{SNR}_{ji} = \sum_{r=1}^{3} (|\rho_{jir}| / \sigma) (\theta_{ji}
%+ (1 -\theta_{ji}) / \sqrt{v_0}) / 3$,
where $r$ indexes the replicate of a drug experiment (with $3$ replicates per
drug).
To assess SNR in a drug case, we pool the signal to noise ratio across all
pathways.  

For GSEA, we called a pathway ``dysregulated'' if the false discovery rate
(FDR) adjusted p-value is less than $5\%$. For the CFA-CAR and EFA models we
identify perturbations based on the centroid estimator of $\theta$ in
Eq.~(\ref{thetahat}) where the threshold $t$ was chosen to control BFDR at
$5\%$ (see Eq.~(\ref{bfdr})).  Restricting the BFDR at $5\%$,
we used a perturbation threshold of $0.72$ for IC20 at 12 hours and $0.50$ for
IC20 at 24 hours for the CFA-CAR model.  For the EFA model, we used thresholds
$0.80$ and $0.70$ for IC20 at 12 hours and IC20 at 24 hours, respectively.

%\begin{figure}[htbp]
%\begin{center}
%\includegraphics[width=0.75\textwidth]{BFDR_thresholds_CFA.pdf}
%\caption{BFDR as a function of the threshold used to call a pathway perturbed
%in the CFA-CAR model. We select the threshold based on an expected BFDR of
%$5\%$. }
%\label{BFDR_thresholds_CFA}
%\end{center}
%\end{figure}
%
%\begin{figure}[htbp]
%\begin{center}
%\includegraphics[width=0.75\textwidth]{BFDR_thresholds_EFA.pdf}
%\caption{BFDR as a function of the threshold used to call a pathway perturbed
%in the EFA model. We select the threshold based on an expected BFDR of $5\%$.}
%\label{BFDR_thresholds_EFA}
%\end{center}
%\end{figure}

\subsection{Assessing model fit}
We assessed model fit by first standardizing the data with respect to its
marginalized likelihood given $\Theta$. If, for sample $i$,
\[
Y_i \given \Lambda, \Phi, \theta_i, \sigma^2, \Psi \sim
N(0, ~~ \Lambda \Phi \Sigma(\theta_i) \Phi^\top \Lambda^\top
+ \sigma^2 \Lambda \Phi \Lambda^\top + \Psi),
\]
then, if $C_i$ denotes the Cholesky factor of the above marginalized covariance
of $Y_i$, we have that $Y_i \sim N(0, C_i C_i^\top)$ and so $\tilde{Y_i}
\cdot C^{-1}Y_{i} \sim N(0, I_p)$, where $I_p$ is the $p$-th order identity
matrix. Thus, to assess model fit we take posterior mean estimates $\hat{C}_i$
of $C_i$, consider standardized gene expressions $\tilde{Y}_i = \hat{C}_i^{-1}
Y_i$ for each sample $i = 1, \ldots, n$, and then check two assumptions:
zero-mean gene expressions and normality.

\begin{figure}[htbp]
\begin{center}
\includegraphics[width=\textwidth]{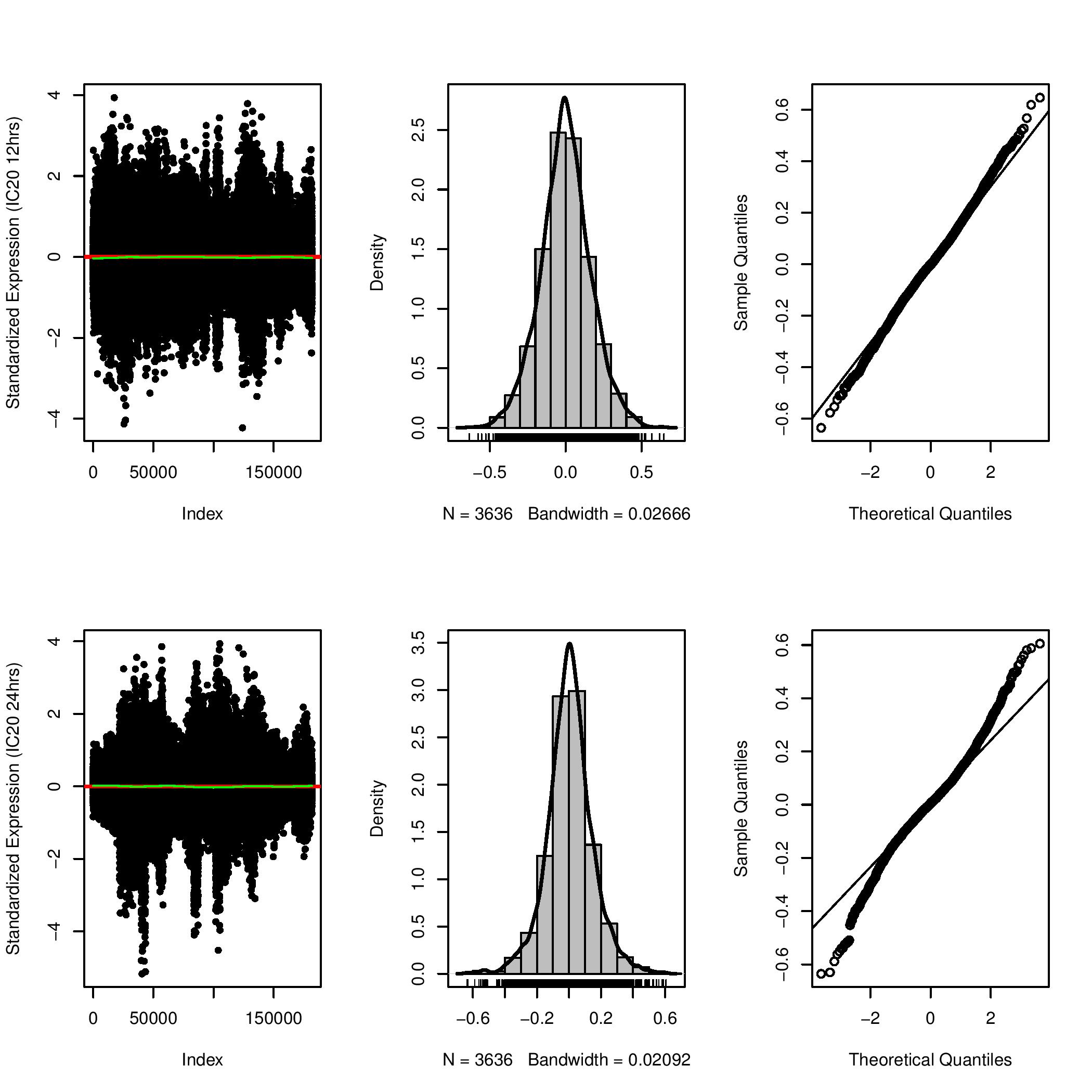}
\caption{Top and bottom panels show model fit assessment for the IC20 at 12
hours and IC20 at 24 hours datasets, respectively.
Left: scatterplot of standardized data with zero expectation and smoother
(lowess) fit shown by a red line and green curve, respectively.
Middle: histogram, density, and rug plot of the standardized dataset averaged
over samples.
Right: quantile-quantile normal plot of standardized data averaged over
samples.
}
\label{modelfit}
\end{center}
\end{figure}

Figure~\ref{modelfit} summarizes our assessment, with samples from IC20 at 12
and 24 hours in top and bottom panels, respectively.
In the first sub-figure in the left, we plot pooled standardized gene
expressions over both samples and genes to assess the zero-mean assumption. As
can be seen from the smoother fit, this assumption is well met, but samples
might have slightly different variance scales. For this reason, since we are
focused on gene expressions, we assess normality via average standardized gene
expressions over samples in the two last sub-figures. We found that the IC20 at 24
hours data has larger departures from normality in the tails when compared to
the IC20 at 12 hours dataset.  This departure at IC20 at 24 hours is specifically due to two of the drugs, H-7 Dihydrochloride (DHCL) and Mitomycin C (MMC), which showed a drastic change in variability between the IC20 at 12 hours and IC20 at 24 hours.  This effect is explained in further detail in Section~\ref{IC2024}.

\subsection{Assessing control of false positives via cross-validation}
\begin{figure}[htbp]
\begin{center}
\includegraphics[width=\textwidth]{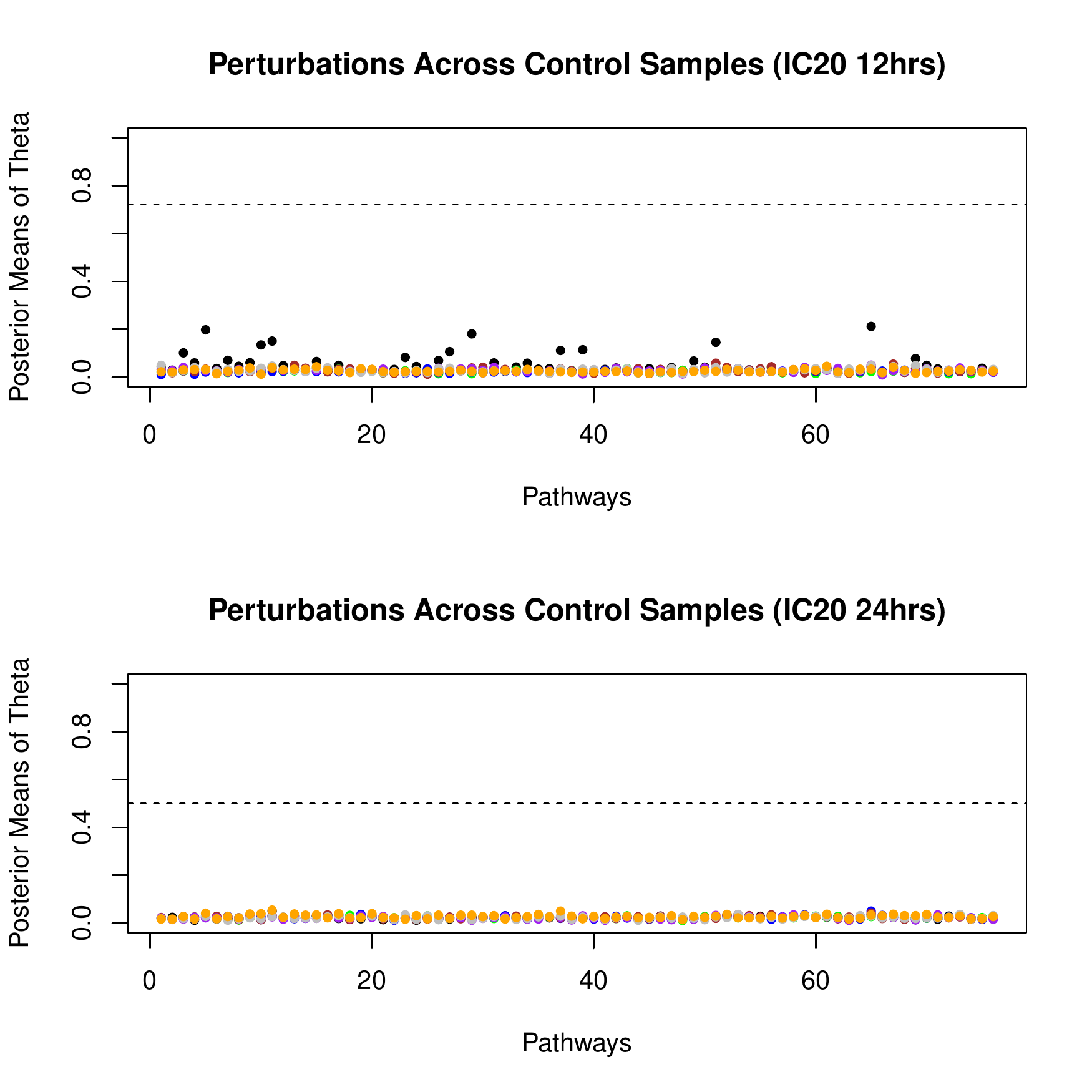}
\caption{Top: Posterior means of $\theta$ across control samples in the IC20
$12$ hours data set (Top) and the IC20 $24$ hours data set (Bottom) in the
leave-one-out cross-validation. Different colors represent different
leave-one-out trials ($8$ trials in total).  The dotted horizontal line
represents the perturbation threshold used to identify perturbations at a BFDR
level of $5\%$.}
\label{cross-validation}
\end{center}
\end{figure}

To verify the accuracy of our method, we tested our model against the control
(mock) data by leaving one mock sample (microarray), out of the control group
and treated it as a ``case'', where the perturbations are left unknown.  We
applied this leave-one-out cross-validation to each of the $8$ control
samples, to verify that the perturbed pathways identified in the case samples
are not similarly identified against a control sample.  We found that in each
leave-one-out cross-validation test, we did not identify any pathways as being
perturbed in the mock sample (Fig.~\ref{cross-validation}). 

\subsection{Comparing EFA and CFA-CAR model results}

Overall, we found that for both data sets, the EFA model was less sensitive
than the CFA model, finding a subset of mechanistically relevant pathways that
were identified by the CFA model.  Because the EFA model identified fewer
pathways linked to drug mechanisms, we choose to focus our comparison with
GSEA  in the following subsections.  Results obtained from the EFA model can
be found in Appendix~\ref{A.EFA_real_data_results}.

\subsection{IC20 concentration at 12 hours}
\label{sec:ic20-12}

\begin{figure}[htpb]
\begin{center}
\includegraphics[width=\textwidth]{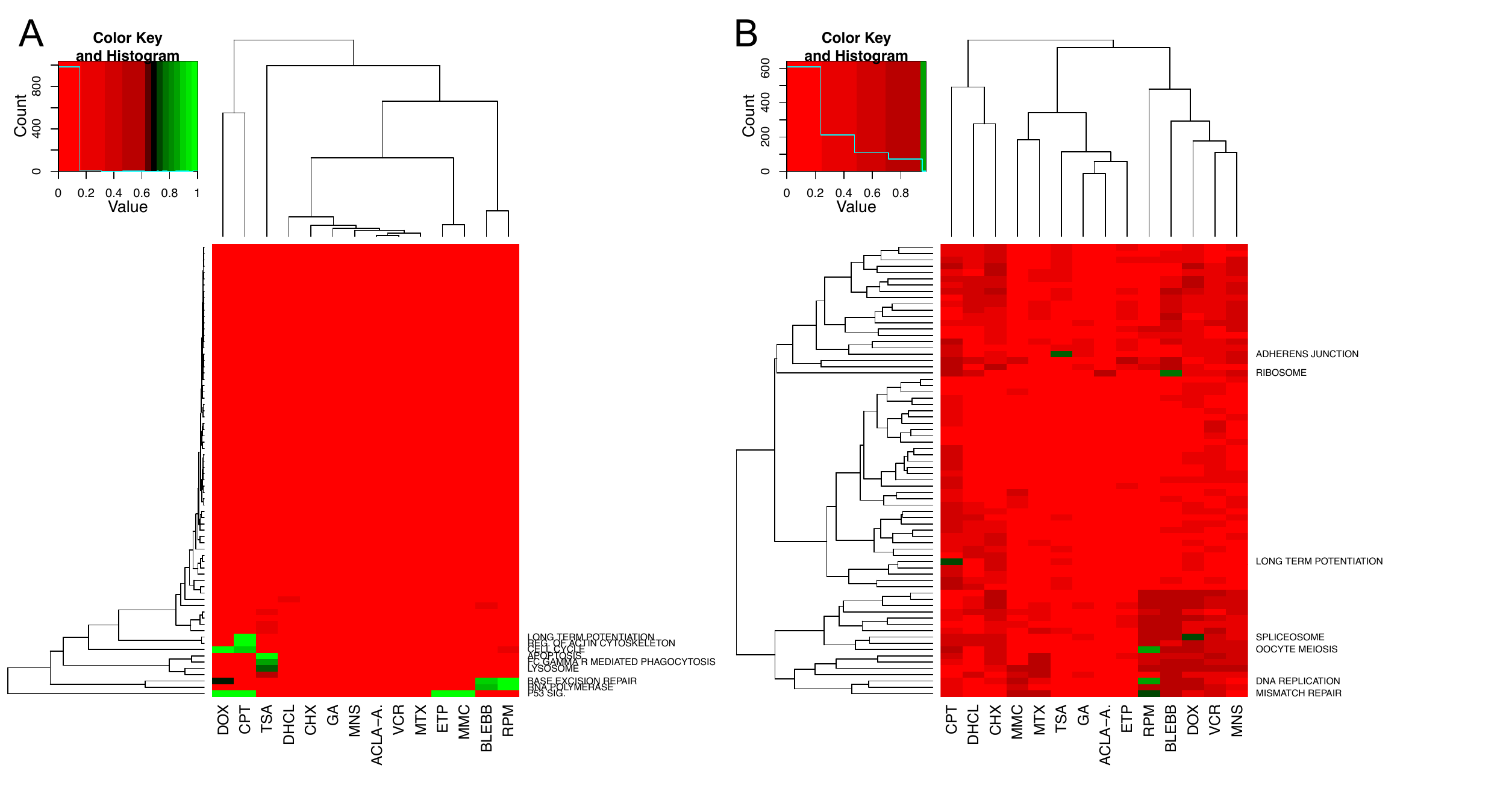}
\caption{A. Heatmap of the CFA-CAR posterior means of $\Theta$ for IC20 at
12 hours.  Green cells indicate high posterior probabilities (above the
$x=0.72$ threshold or equivalently, less than $5\%$ BFDR).
B. Heatmap of (1-FDR adjusted P-values from) GSEA for
IC20 at 12 hours.  Green cells indicate small FDR values (above a $5\%$
threshold).
In both heatmaps, x-axes indicate drug experiments (samples) and y-axes
indicate pathways. Each axis is clustered hierarchically using complete
linkage and Euclidean distances as dissimilarities.}
\label{heatmap_IC20_12hrs}
\end{center}
\end{figure}

For the CFA-CAR model results, we ran a hierarchical bi-clustering analysis on
the posterior means $\hat{\Theta}$ of $\Theta$. More specifically, we
independently cluster samples (drugs) and pathways using complete linkage and
Euclidean distances as dissimilarity metrics. Each dimension has their
observations swapped to match the clustering hierarchy, as can be seen in
Fig~\ref{heatmap_IC20_12hrs}A. Fig.~\ref{heatmap_IC20_12hrs}B shows the
results from a similar analysis on adjusted p-values obtained from GSEA for
comparison.

At IC20 at 12 hours, the most commonly identified pathway across drugs by the
CFA-CAR model was P53-signaling: it was identified for etoposide (ETP),
mitomycin-C (MMC), camptothecin (CPT), and doxorubicin (DOX). 
Interestingly, all four of these compounds are DNA damaging agents and were
clustered into two groups (ETP/MMC and CPT/DOX) (Fig.~\ref{heatmap_IC20_12hrs}),
whereas some of the compounds that affect protein function unrelated to DNA
damage, such as geldanamycin (GA) and rapamycin (RPM), also formed their own
cluster.  P53 is an important tumor suppressor that can regulate various
cellular processes including apoptosis, cell cycle, and DNA repair;
furthermore, it is mutated in over $50\%$ of human cancers
\citep[reviewed in][]{ling_review}.  This loss of P53-signaling can lead to
cancer growth and propagation.  In many cases, P53-signaling is induced as a
response to specific DNA damaging agents \citep{gupta}, playing key roles in
cell cycle arrest and apoptosis.  All four drugs have been shown to play
significant roles in activating P53-dependent mechanisms
[ETP:\@\citep{karpinich,grandela},
MMC:\@\citep{abbas,fritsche,verweij},
CPT:\@\citep{jaks,gupta,wang},
DOX:\@\citep{kurz,ling,zhou}].
Furthermore, for some of these compounds including CPT and DOX, cell cycle
arrest at specific check points occurs through the induction of a P53
signaling mechanism.  For instance, P53 signaling was shown to play a critical
role in the G1 checkpoint of the cell cycle under CPT-induced DNA damage
\citep{jaks,gupta}.  DOX is another chemotherapeautic DNA-damaging drug that
causes cell cycle arrest in the G2/M phase \citep{ling}.  Furthermore, DOX was
shown to cause an accumulation of P53 that lead to a depletion of cells in the
G2/M phase and apoptosis \citep{zhou}.  

GSEA did not identify P53-signaling or cell cycle for any of the 14
compounds.  Morever, the connection between the set of pathways identified by
GSEA and some of the drug's primary MoAs were not readily apparent.  Both GSEA
and the CFA-CAR model identified DNA repair pathways for Rapamycin (RPM), with
GSEA identifying mismatch repair and CFA-CAR identifying base excision repair,
although the connection between these repair pathways and RPM's MoA is unclear.

\subsection{IC20 concentration at 24 hours}\label{IC2024}
\begin{figure}[htpb]
\begin{center}
\includegraphics[width=\textwidth]{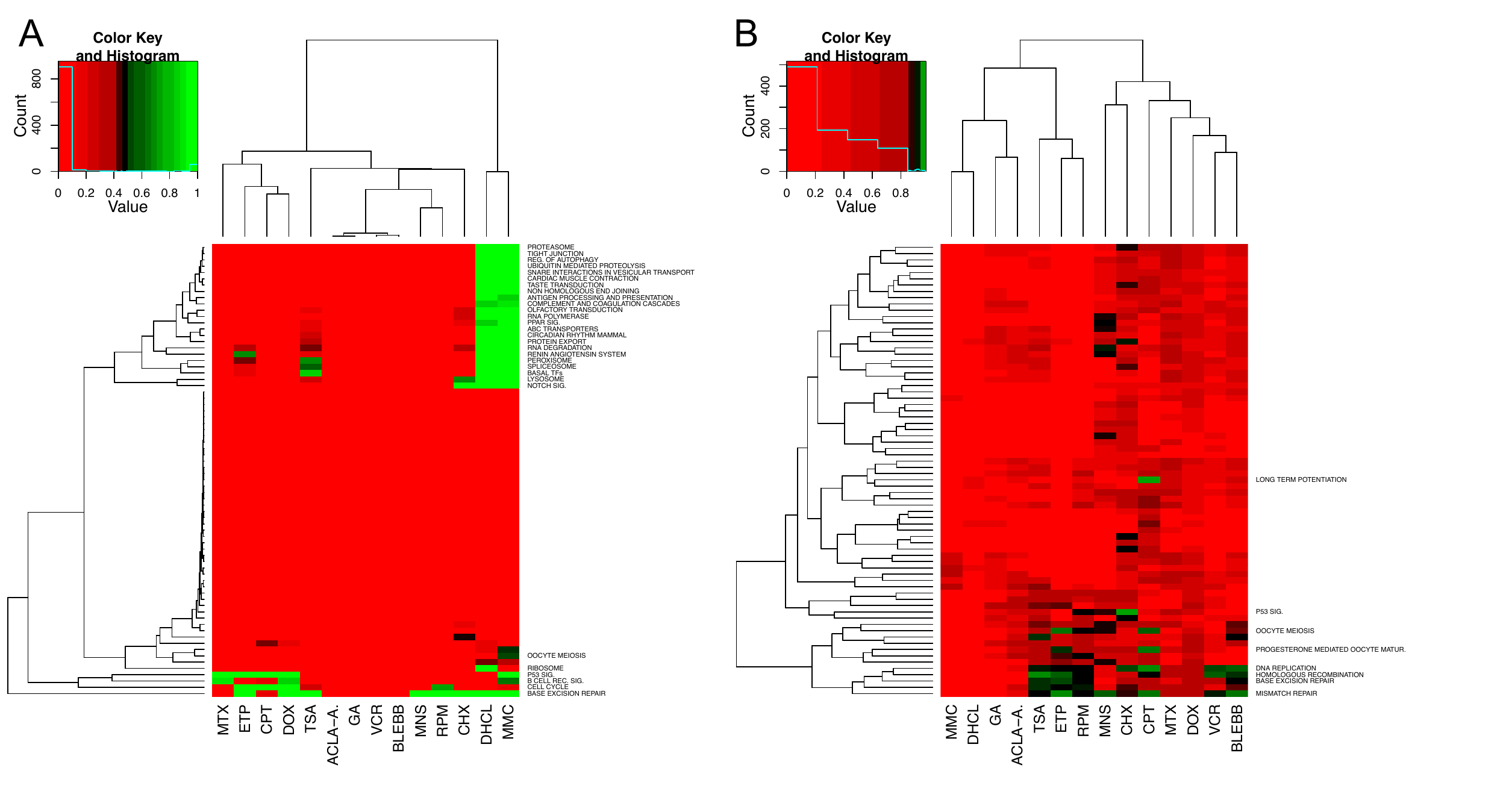}
\caption{A. Heatmap of the CFA-CAR posterior means of $\Theta$ for IC20 at
24 hours.  Green cells indicate high posterior probabilities (above the
$x=0.50$ threshold or equivalently, less than $5\%$ BFDR).
B. Heatmap of (1-FDR adjusted P-values from) GSEA for
IC20 at 24 hours.  Green cells indicate small FDR values (above a $5\%$
threshold).
In both heatmaps, x-axes indicate drug experiments (samples) and y-axes
indicate pathways. Each axis is clustered hierarchically using complete
linkage and Euclidean distances as dissimilarities.}
\label{IC20_24hrs_heatmap}
\end{center}
\end{figure}

At longer exposure times, both methods picked up more pathways (see
Fig.~\ref{IC20_24hrs_heatmap} for heatmaps of the CFA-CAR and GSEA results,
respectively.) This is expected from a biological point of view because as
time of exposure increases the cell needs to respond to potentially more
complex effects caused by the loss of cell homeostasis. Therefore, cells
activate more pathways that, for example, help in cell survival, increase
energy output, or signal cell demise or death.
With the CFA-CAR model, we found that at IC20 concentrations
at 24 hours of exposure, H-7 Dihydrochloride (DHCL) and Mitomycin C (MMC)
were overly perturbed (i.e.\ many pathways had a significantly high posterior
mean probability of having a non-zero perturbation).  Base excision repair was
again identified for RPM but was also picked by several other drugs such as
monastrol (MNS), cycloheximide (CHX), trichostatin-A (TSA), DOX, and ETP\@.
As mentioned earlier, the connection between DNA repair pathways and these
compounds is unclear.  Moreover, GSEA identified various DNA damaging repair
or DNA related pathways at an FDR level of $5\%$.  Such pathways include base
excision repair, homologous recombination, mismatch repair, and DNA replication. 

P53-signaling remains one of the most perturbed pathways among several of
DNA-damaging or inhibiting compounds at the longer exposure time, including
ETP, CPT, and DOX, all of which clustered together. P53-signaling was also
picked up for methotrexate (MTX) which was something that was not picked up at
the shorter exposure time but is very much linked to P53 induction
\citep{krause,wen}.  \citet{krause} showed that treating HepG2 cells with
methotrexate increases levels of p53.  \citet{wen} also demonstrated that
methotrexate can induce apoptosis by the induction of the P53 targeted genes,
DR5, P21, Puma and Noxa.  Similarly, GSEA also identified P53 signaling for
CHX, an inhibitor of protein biosynthesis at the longer exposure time, but was
missed by CFA-CAR\@. Cell cycle was picked up for DOX and CPT (as in the
shorter exposure time), but under these stronger experimental conditions, cell
cycle was further identified for ETP and RPM as well.

\section{Discussion} 
\label{sec:discussion}

Identifying mechanisms of action from transcriptional data alone is
challenging.  In disease phenotypes, differences in gene expression cannot, by
themselves, elucidate aberrant causal pathways.  Drug perturbations are even
more difficult because often times perturbations occur directly on a proteomic
level rather than a gene level.  In many cases, transcriptional responses are
the residual aftermath of a perturbation rather than being representative of
the direct target. 

GSEA and other gene set methods attempt to boost the signal on transcriptional
data by focusing on entire gene sets with the assumption that related genes
may contribute a greater signal than individual genes alone.  Furthermore,
network-based methods use cellular regulation or gene/protein interaction data
to better understand the MoAs underlying a transcriptional response.

Here, we link transcriptional data to unobserved mechanisms of action using
confirmatory factor analysis in conjunction with a conditional autoregressive
model.  In the CFA level of the model we link gene expression to the
biological pathways that contain their gene products.  In the CAR model, we
seek to explain the variation observed in these latent pathways  by a
pathway-pathway network induced by ``external'' perturbations.  

The core goal of this model is to uncover these perturbation targets.  We have
shown through simulation and real data that providing a network filtering
method on expression data is a significant improvement over non-network
approaches. 
In simulation, we show that the CFA-CAR model exhibits a high specificity
across various signal to noise ratios, unlike an EFA model  (network-free)
which has a higher false positive rate as the signal in the data set
increases.  Furthermore, we have shown that the CFA-CAR model is very robust
to inaccuracies in the databases used to create the pathways and the
pathway-pathway interaction network.

Moreover, our method was able to identify pathways significantly related to
some of the drugs' MoAs in the NCI DREAM 7 data sets.  Specifically, we were
able to identify signaling/regulatory pathways that play a causative role for
many of the DNA-damaging compounds.  These pathways were generally not
identified using gene set enrichment analysis, or less so with EFA\@. Some of
the known targets were missed by both the CFA-CAR model and GSEA across both
data sets.  For instance, rapamycin targets the mTOR protein \citep{alqurashi}
but mTOR signaling was not identified.  In another example, blebbistatin
inhibits the myosin II protein \citep{allingham}, and we expected to see some
of the cellular motility pathways such as regulation of cytoskeleton pathway.
Moreover, some of the results we found with both methods were also unclear
(e.g.\ identifying DNA repair pathways across many of these compounds).  

There are several natural directions for extension of this model.  For
instance, other forms of biological variables such as transcription factors or
metabolic pathways could be used as latent factors.  Another extension of this
model would be to integrate other forms of high-throughput data in conjunction
with gene expression. A temporal component could also be incorporated to
accommodate temporal data and define a dynamic model. All these extensions,
however, are at the cost of increased model complexity, which can greatly
affect the fit and convergence of the model.

\appendix

\makeatletter   %% HAVE TO ADD SOMETHING HERE TO MAKE IT SAY "APPENDIX"
 \renewcommand{\@seccntformat}[1]{APPENDIX~{\csname the#1\endcsname}.\hspace*{1em}}
 \makeatother

\section{Network Construction}
\label{A.network_construction}

Our network $W$ was created in a similar manner to \citet{lpia}.  We constructed a weighted network of pathways $W$ where the nodes represent
canonical pathways and the edges represent functional links.  We describe the
details of the construction of $W$ below.

First, we classified genes using two sources of biological information: 
KEGG regulatory/signaling pathways \citep{kegg2,kegg3}, and GO
biological processes \citep{go} obtained from the MSigDB collection
\citep{gsea} (version 3.0).  We removed disease specific pathways and any
metabolic pathways, focusing our attention to a regulatory/signaling network.
We also removed 4 pathways that did not represent a single pathway but a
collection of signaling molecules that are themselves part of other pathways
(e.g. Cell Adhesion Molecules and Cytokine-Cytokine Receptor Interactions).
In the end, we had 72 distinct regulatory/signaling KEGG pathways.

To define functional links between two pathways, we constructed a bipartite
network, where the two sets of nodes represent KEGG pathways and GO functions,
respectively.  A pathway $P$ and a GO term $G$ were linked together in this
bipartite network if the intersection of $P$ and $G$ (as gene sets) is
non-empty. Furthermore, the edge was weighted by the Jaccard index between $G$
and $P$, measuring the relative overlap between function and
pathway. Edges with Jaccard index smaller than $3\%$ were removed to reduce
the possibility of false positives in the database.  We can represent this
bipartite graph as an incidence matrix $M$ where the rows represent KEGG
pathways and the columns represent GO terms.

Our final network of pathways was obtained by translating the information from
this two-mode network onto a one mode network formed by the KEGG node set
alone.   The projection of $M$ onto a single mode network $A$ was obtained by
$A = M^\top M$. As a result, two pathways are linked in this network if and
only if they share at least one biological process.  Furthermore, edges
between pathways that heavily contribute to the same GO terms will be heavier
than those pathways that contribute less to the same GO terms.
Finally, we standardized $A$ in the following manner:
\[
W_{ij} = \frac{A_{ij} }{\sqrt{A_{ii} A_{jj}} } (1 - \delta_{ij}),
\]
where $\delta$ is the Kronecker delta.

\section{Marginalized conditional posterior distribution of $\theta$ }
\label{A.theta}

In this section, we derive the marginalized posterior of $\theta_i$ for sample
$i$.  After integrating out $\rho_i$ from Eq.~(\ref{omega_prior}) we have that
\[
\omega_i \given \theta_i, \Phi \ind N(0, V(\theta_i)),
\]
where $V(\theta_i) = \sigma^2 \Phi + \Phi \Sigma(\theta_i) \Phi^\top$ and the
index runs over samples. To simplify the notation, we set $\tau_0^2 := v_0
\tau^2$ and drop the indices, that is,
$\omega \given \theta, \Phi \sim N(0, V(\theta))$.

Since $\theta_j \iid \text{\sf Bern}(\alpha)$ from the prior and
with $\Sigma(\theta)$ similarly defined as in Section~\ref{rho_posterior}, we
have that
\begin{align*}
V(\theta) &= \sigma^2 \Phi + \tau_0^2 \Phi\Phi^\top
+ (\tau^2-\tau_0^2) \Phi \text{\sf Diag}(\theta) \Phi^\top \\
&= \sigma^2 \Phi + \tau_0^2 \Phi\Phi^\top
+ (\tau^2 - \tau_0^2) \sum_{j=1}^q \theta_j \phi_j \phi_{j}^\top,
\end{align*}
where $\phi_j$ is the $j$-th column of $\Phi$.

We want
\[
\Pr(\theta_j = 1 | \theta_{[-j]}, \omega, \Phi) =
\frac{\Pr(\theta_j = 1, \theta_{[-j]}, \omega, \Phi)}
{\sum_{b \in \{0,1\}} \Pr(\theta_j = b, \theta_{[-j]}, \omega, \Phi)}.
\]
By the distributions above,
\begin{multline*}
\Pr(\theta_j, \theta_{[-j]}, \omega, \Phi) \propto
\prod_j \alpha^{\theta_j} {(1-\alpha)}^{1-\theta_j} \\
{(2\pi)}^{-q/2} {|V(\theta)|}^{-1}
\exp\Bigg\{-\frac{1}{2} \omega^\top {V(\theta)}^{-1} \omega\Bigg\}. \\
\end{multline*}

Now, we define $V_0 = V(\theta_j=0, \theta_{[-j]})$ and
$V_1 = V(\theta_j=1, \theta_{[-j]})$. Then,
\begin{align*}
V_1 &= \sigma^2 \Phi + \tau_0^2 \Phi\Phi^\top
+ (\tau^2-\tau_0^2) \phi_j \phi_j^\top \\
& \quad + (\tau^2-\tau_0^2) \sum_{k \ne j} \theta_k \phi_k \phi_k^\top \\
&= V_0 + (\tau^2-\tau_0^2) \phi_j \phi_j^\top,
\end{align*}
and so, by the Sherman-Morrison formula,
\begin{align*}
V_1^{-1} &= {(V_0 + (\tau^2 - \tau_0^2) \phi_j \phi_j^\top)}^{-1} \\
&= V_0^{-1} - \frac{(\tau^2 - \tau_0^2) V_0^{-1} \phi_j \phi_j^\top V_0^{-1}}
{1 + (\tau^2 - \tau_0^2)\phi_j^\top V_0^{-1} \phi_j}.
\end{align*}
Moreover, by the matrix determinant lemma, 
\[
|V_1| = (1 + (\tau^2 - \tau_0^2) \phi_j^\top V_0^{-1} \phi_j)|V_0|.
\]

Using these relations we can proceed to sample $\theta_j^{(t+1)}$ based on
$\theta^{(t)}$ from the current iteration $t$; we keep the inverse of
$V(\theta)$ and only update it when $\theta_j$ is flipped.  We have
\begin{multline*}
\Pr(\theta_j^{(t+1)} = 1 \given \theta_{[-j]}, \omega, \Phi) = \\
\frac{|V_1|^{-1/2} \exp\{-\frac{\omega^\top V_1^{-1} \omega}{2}\} \alpha}
{|V_1|^{-1/2} \exp\{-\frac{\omega^\top V_1^{-1} \omega}{2}\} \alpha
+ |V_0|^{-1/2} \exp\{-\frac{\omega^\top V_0^{-1} \omega}{2}\} (1 - \alpha)},
\end{multline*}
or, taking logits,
\begin{equation}
\label{apx:logp}
\logit \Pr\Big(\theta_j^{(t+1)} = 1 | \theta_{[-j]}, \omega, \Phi\Big) =
-\frac{1}{2}\log\frac{|V_1|}{|V_0|} 
-\frac{1}{2} \omega^\top (V_1^{-1}-V_0^{-1}) \omega + \logit(\alpha).
\end{equation}

There are two cases: when $\theta_j^{(t)} = 0$ and when $\theta_j^{(t)} = 1$.
In the first case we already have $V_0^{-1}$, and so, by defining
$\delta_{j0} = 1 + (\tau^2 -\tau_0^2) \phi_j^\top V_0^{-1} \phi_j$ and
$\Delta_{j0} = (\tau^2 - \tau_0^2) /
\delta_{j0} (V_0^{-1}\phi_j){(V_0^{-1}\phi_j)}^\top$ we can apply
Eq.~(\ref{apx:logp}) to obtain
%
%\begin{multline*}
\[
\logit\, \Pr\Big(\theta_j^{(t+1)} = 1 \given \theta_{[-j]}, \omega, \Phi\Big)
= %\\
-\frac{1}{2} \log \delta_{j0}
+ \frac{1}{2} \omega^\top \Delta_{j0} \omega
+ \logit(\alpha).
\]
%\end{multline*}
%
In the sampler, we update ${({V(\theta)}^{-1})}^{(t+1)} =
{({V(\theta)}^{-1})}^{(t)} - \Delta_{j0}$ if ${\theta_j}^{(t+1)} = 1$, that
is, when $\theta_j$ is flipped.

If $\theta_j^{(t)} = 1$ we define, similarly,
$\delta_{j1} = 1 - (\tau^2 - \tau_0^2) \phi_j^\top V_1^{-1} \phi_j$ and
$\Delta_{j1} = (\tau^2 - \tau_0^2)/\delta_{j1} (V_1^{-1} \phi_j)
{(V_1^{-1} \phi_j)}^\top$ to obtain:
%
%\begin{multline*}
\[
\logit\, \Pr\Big(\theta_j^{(t+1)} = 1 \given \theta_{[-j]}, \omega, \Phi\Big)
= %\\
\frac{1}{2}\log\delta_{j1}
+ \frac{1}{2}\omega^\top \Delta_{j1} \omega
+ \logit(\alpha),
\]
%\end{multline*}
%
and update
${({V(\theta)}^{-1})}^{(t+1)} = {({V(\theta)}^{-1})}^{(t)} + \Delta_{j1}$
if $\theta_j$ gets flipped to $\theta_j^{(t+1)} = 0$.

\section{EFA Results on the DREAM 7 Data Sets}
\label{A.EFA_real_data_results}

\begin{figure}[htpb]
\begin{center}
\includegraphics[width=\textwidth]{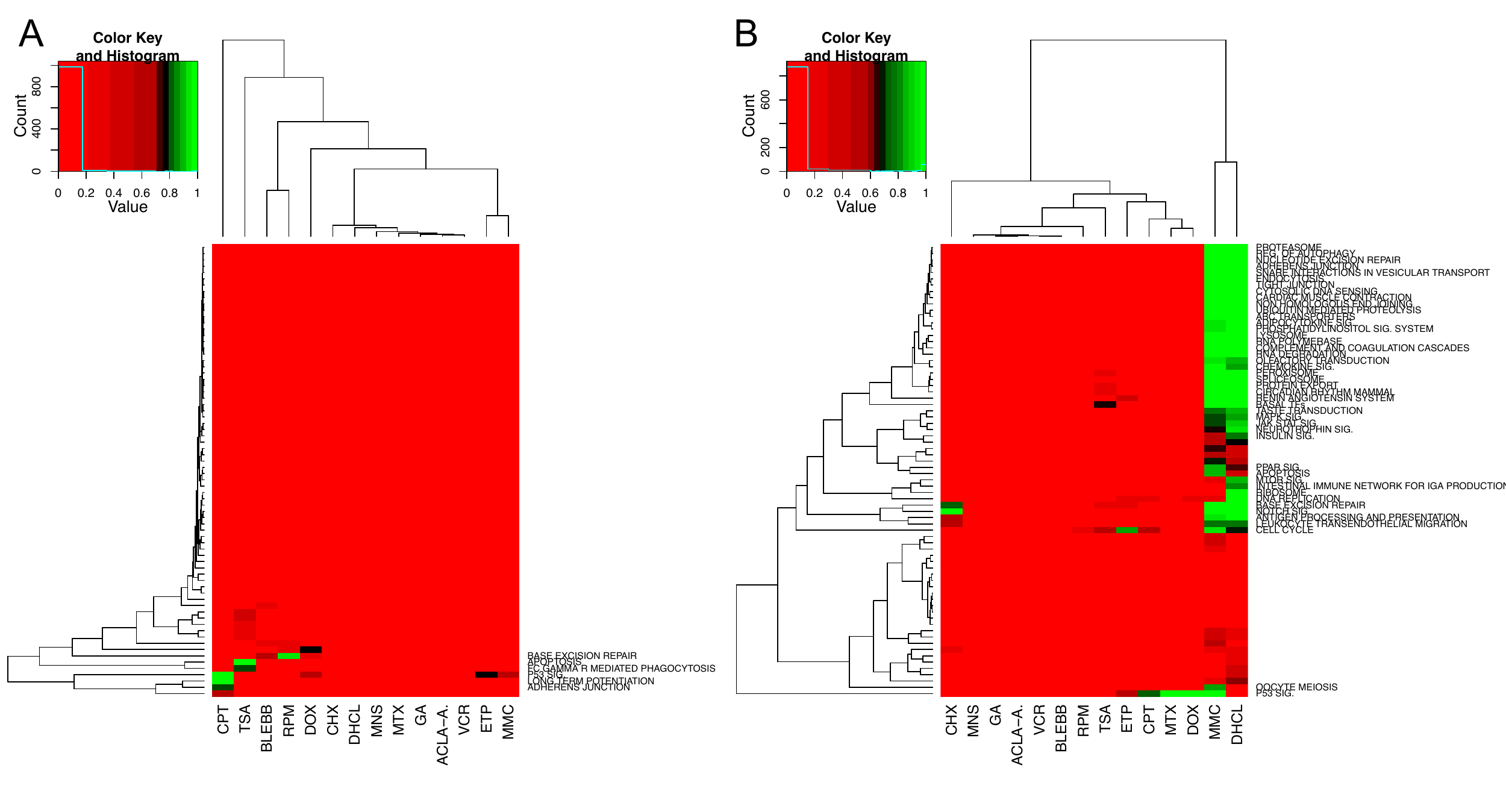}
\caption{A. Heatmap of the EFA posterior means of $\Theta$ for IC20 at
12 hours.  Green cells indicate high posterior probabilities (above the
$x=0.80$ threshold or equivalently, less than $5\%$ BFDR).
B. Heatmap of the EFA posterior means of $\Theta$ for IC20 at
24 hours.  Green cells indicate high posterior probabilities (above the
$x=0.70$ threshold or equivalently, less than $5\%$ BFDR).
In both heatmaps, x-axes indicate drug experiments (samples) and y-axes
indicate pathways. Each axis is clustered hierarchically using complete
linkage and Euclidean distances as dissimilarities.}
\label{heatmap_IC20_12hrs_and_24hrs_EFA}
\end{center}
\end{figure}

Fig~\ref{heatmap_IC20_12hrs_and_24hrs_EFA} includes heatmaps of the results
obtained by the EFA model for the IC20 at 12 hours and IC20 at 24 hours
exposure times, respectively.

At IC20 12 hours, the EFA model identified fewer pathways than the CFA-CAR
model, identifying P53-signaling for CPT (also identified by CFA-CAR in
addition to other DNA-damaging agents ETP, MMC, and DOX). At IC20 24
hours, the EFA and CFA-CAR models both identified P53-signaling for the same
pathways (MMC, DOX, MTX, and CPT), as well as cell cycle for ETP, a pathway
mechanistically linked to p53-signaling.  However, CFA-CAR further identified
cell cycle for CPT, DOX, and RPM\@.  Moreover, both EFA and CFA-CAR both found
MMC and DHCL overly perturbed at IC20 24 hours.

%% HERE WE DECLARE THE BIBLIOGRAPHYSTYLE TO USE AND THE BIBLIOGRAPHY DATABASE
%\bibliographystyle{ECA_jasa}
\bibliographystyle{asa}
\if0\blind{
\bibliography{cfa-car_model_08272014}
}
\else{
\bibliography{JASA_blind_cfa-car}

\begin{thebibliography}{74}
\newcommand{\enquote}[1]{``#1''}
\expandafter\ifx\csname natexlab\endcsname\relax\def\natexlab#1{#1}\fi

\bibitem[{Abbas et~al.(2002)Abbas, Olivier, Lopez, Houser, Xiao, Suresh-Kumar,
  Tomasz, and Bargonetti}]{abbas}
Abbas, T., Olivier, M., Lopez, J., Houser, S., Xiao, G., Suresh-Kumar, G.,
  Tomasz, M., and Bargonetti, J. (2002), \enquote{Differential activation of
  p53 by the various adducts of mitomycin {C},} \textit{Biol Chem.}, 277,
  40513–--40519.

\bibitem[{Allingham et~al.(2005)Allingham, Smith, and Rayment}]{allingham}
Allingham, J.~S., Smith, R., and Rayment, I. (2005), \enquote{The structural
  basis of blebbistatin inhibition and specificity for myosin {II},}
  \textit{Nature Structural \& Molecular Biology}, 12, 378--379.

\bibitem[{Alqurashi et~al.(2013)Alqurashi, Hashimi, and Wei}]{alqurashi}
Alqurashi, N., Hashimi, S.~M., and Wei, M.~Q. (2013), \enquote{Chemical
  Inhibitors and micro{RNA}s (mi{RNA}) Targeting the Mammalian Target of
  Rapamycin (m{TOR}) Pathway: Potential for Novel Anticancer Therapeutics,}
  \textit{Int J Mol Sci.}, 14, 3874--3900.

\bibitem[{Bansal et~al.(2014)Bansal, Yang, Karan, Menden, Costello, Tang, Xiao,
  Li, Allen, Zhong, Chen, Kim, Wang, Heiser, Realubit, Mattioli, Alvarez, Shen,
  community, Gallahan, Singer, Saez-Rodriguez, Xie, Stolovitzky, and
  Califano}]{bansal}
Bansal, M., Yang, J., Karan, C., Menden, M.~P., Costello, J.~C., Tang, H.,
  Xiao, G., Li, Y., Allen, J., Zhong, R., Chen, B., Kim, M., Wang, T., Heiser,
  L., Realubit, R., Mattioli, M., Alvarez, M., Shen, Y., community, N.-D.,
  Gallahan, D., Singer, D., Saez-Rodriguez, J., Xie, Y., Stolovitzky, G., and
  Califano, A. (2014), \enquote{The challenge of predicting synergistic and
  antagonistic compound-pair activity from individual compound perturbations,}
  \textit{Nature Biotechnology}, in-press.

\bibitem[{Bollen(1989)}]{bollen}
Bollen, K.~A. (1989), \textit{Structural Equations with Latent Variables}, New
  Jersey: Wiley-Interscience.

\bibitem[{Braunstein et~al.(2011)Braunstein, McShane, Piette, and
  Jensen}]{piette}
Braunstein, A., McShane, B., Piette, J., and Jensen, S. (2011), \enquote{A
  Hierarchical {B}ayesian Variable Selection Approach to Major League Baseball
  Hitting Metrics,} \textit{JQAS}, 7.

\bibitem[{Carvalho et~al.(2008)Carvalho, Chang, Lucas, Wang, Nevins, and
  West}]{west_2008}
Carvalho, C., Chang, J., Lucas, J., Wang, Q., Nevins, J., and West, M. (2008),
  \enquote{High-dimensional sparse factor modelling: Applications in gene
  expression genomics,} \textit{J. Am. Stat. Assoc.}, 103, 1438--–1456.

\bibitem[{Carvalho and Lawrence(2008)}]{carvalho}
Carvalho, L.~E. and Lawrence, C.~E. (2008), \enquote{Centroid estimation in
  discrete high-dimensional spaces with applications in biology,}
  \textit{Proceedings of the National Academy of Sciences}, 105, 3209--3214.

\bibitem[{Chuang et~al.(2007)Chuang, Lee, Lee, and Ideker}]{ideker_2007}
Chuang, H.~Y., Lee, E.and~Liu, Y.~T., Lee, D., and Ideker, T. (2007),
  \enquote{Network-based classification of breast cancer metastasis,}
  \textit{Molecular Systems Biology}, 3, 140.

\bibitem[{Cosgrove et~al.(2008)Cosgrove, Yingchun, Gardner, and
  Kolaczyk}]{cosgrove_2008}
Cosgrove, E.~J., Yingchun, Z., Gardner, T.~S., and Kolaczyk, E.~D. (2008),
  \enquote{Predicting gene targets of perturbations via network-based filtering
  of mRNA expression compendia,} \textit{BMC Bioinformatics}, 24, 2482--2490.

\bibitem[{Cressie(1993)}]{cressie}
Cressie, N. A.~C. (1993), \textit{Statistics for Spatial Data}, New York:
  Wiley.

\bibitem[{de~Oliveira(2012)}]{oliveira}
de~Oliveira, V. (2012), \enquote{Bayesian Analysis of Conditional
  Autoregressive Models,} \textit{Ann Inst Stat Math.}, 64, 107--133.

\bibitem[{di~Bernardo et~al.(2005)di~Bernardo, Thompson, Gardner,
  et~al.}]{bernardo_2005}
di~Bernardo, D., Thompson, M., Gardner, T., et~al. (2005),
  \enquote{Chemogenomic profiling on a genome-wide scale using
  reverse-engineered gene networks,} \textit{Nature Biotechnology}, 23,
  377--383.

\bibitem[{Draghici et~al.(2007)Draghici, Khatri, Tarca, Amin, Done, Voichita,
  Georgescu, and Romero}]{pathwayexpress}
Draghici, S., Khatri, P., Tarca, A.~L., Amin, K., Done, A., Voichita, C.,
  Georgescu, C., and Romero, R. (2007), \enquote{A systems biology approach for
  pathway level analysis,} \textit{Genome Res}, 17, 1537--1545.

\bibitem[{Dudoit et~al.(2003)Dudoit, Shaffer, and Boldrick}]{Dudoit}
Dudoit, S., Shaffer, J.~P., and Boldrick, J.~C. (2003), \enquote{Multiple
  hypothesis testing in microarray experiments,} \textit{Stat. Science}, 18,
  71--103.

\bibitem[{Faulon et~al.(2008)Faulon, Misra, Martin, Sale, and Sapra}]{faulon}
Faulon, J.~L., Misra, M., Martin, S., Sale, K., and Sapra, R. (2008),
  \enquote{Genome scale enzyme-metabolite and drug-target interaction
  predictions sing the signature molecular descriptor,}
  \textit{Bioinformatics}, 24, 225--233.

\bibitem[{Figueiredo-Pereira~M et~al.(1996)Figueiredo-Pereira~M, Chen, Li, and
  Johdo}]{acla}
Figueiredo-Pereira~M, E., Chen, W.~E., Li, J., and Johdo, O. (1996),
  \enquote{The antitumor drug aclacinomycin A, which inhibits the degradation
  of ubiquitinated proteins, shows selectivity for the chymotrypsin-like
  activity of the bovine pituitary 20 S proteasome,} \textit{Journal of
  Biological Chemistry}, 271, 16455--9.

\bibitem[{Fritsche et~al.(1993)Fritsche, Haessler, and Brandner}]{fritsche}
Fritsche, M., Haessler, C., and Brandner, G. (1993), \enquote{Induction of
  nuclear accumulation of the tumor-suppressor protein p53 by {DNA} damaging
  agents,} \textit{Oncogene}, 8, 307--–318.

\bibitem[{Gandhi et~al.(2006)Gandhi, Zhong, Mathivanan, et~al.}]{gandhi_2006}
Gandhi, T., Zhong, J., Mathivanan, S., et~al. (2006), \enquote{Analysis of the
  human protein interactome and comparison with yeast, worm and fly interaction
  datasets,} \textit{Nature genetics}, 38, 285--293.

\bibitem[{Gelman(2006)}]{gelman_2006}
Gelman, A. (2006), \enquote{Prior distributions for variance parameters in
  hierarchical models,} \textit{Bayesian Analysis}, 1, 515--533.

\bibitem[{Gelman et~al.(2004)Gelman, Carlin, Stern, and Rubin}]{gelman}
Gelman, A., Carlin, J.~B., Stern, H.~S., and Rubin, D.~B. (2004),
  \textit{Bayesian Data Analysis}, New York: Chapman \& Hall/CRC.

\bibitem[{George and McCulloch(1997)}]{george}
George, E.~I. and McCulloch, R.~E. (1997), \enquote{Approaches for {B}ayesian
  variable selection,} \textit{Statistica Sinica}, 7, 339–--373.

\bibitem[{Goodsell(1999)}]{goodsell}
Goodsell, D.~S. (1999), \enquote{The Molecular Perspective: Methotrexate,}
  \textit{The Oncologist}, 4, 340--341.

\bibitem[{Grandela et~al.(2007)Grandela, Pera, Grimmond, Kolle, and
  Wolvetang}]{grandela}
Grandela, C., Pera, M.~F., Grimmond, S.~M., Kolle, G., and Wolvetang, E.~J.
  (2007), \enquote{p53 is required for etoposide-induced apoptosis of human
  embryonic stem cells,} \textit{Stem Cell Res.}, 1, 116--128.

\bibitem[{Grenert et~al.(1997)}]{grenert}
Grenert, J.~P. et~al. (1997), \enquote{The amino-terminal domain of heat shock
  protein 90 (hsp90) that binds geldanamycin is an {ATP/ADP} switch domain that
  regulates hsp90 conformation,} \textit{J Biol Chem}, 272, 23843--23850.

\bibitem[{Gu et~al.(2010)Gu, Chen, Li, and Li}]{Gu}
Gu, J., Chen, Y., Li, S., and Li, Y. (2010), \enquote{Identification of
  responsive gene modules by network-based gene clustering and extending:
  application to inflammation and angiogenesis,} \textit{BMC Sys Bio}, 4, 47.

\bibitem[{Gupta et~al.(1997)Gupta, Fan, Zhan, Kohn, O'Connor, and
  Pommier}]{gupta}
Gupta, M., Fan, S., Zhan, Q., Kohn, K.~W., O'Connor, P.~M., and Pommier, Y.
  (1997), \enquote{Inactivation of p53 increases the cytotoxicity of
  camptothecin in human colon {HCT116} and breast {MCF-7} cancer cells,}
  \textit{Clin Cancer Res.}, 3, 1653--1660.

\bibitem[{Hidaka et~al.(1984)}]{dhcl}
Hidaka et~al. (1984), \enquote{Isoquinolinesulfonamides, novel and potent
  inhibitors of cyclic nucleotide dependent protein kinase and protein kinase
  C,} \textit{Biochemistry}, 23, 5036.

\bibitem[{Huang et~al.(2011)Huang, Yang, Chang, Marquez, and Chen}]{wen}
Huang, W.~Y., Yang, P.~M., Chang, Y.~F., Marquez, V.~E., and Chen, C.~C.
  (2011), \enquote{Methotrexate induces apoptosis through p53/p21-dependent
  pathway and increases {E}-cadherin expression through downregulation of
  {HDAC/EZH2},} \textit{Biochem Pharmacol.}, 81, 510--517.

\bibitem[{Ideker et~al.(2002)Ideker, Ozier, and Schwikowski}]{ideker_2002}
Ideker, T., Ozier, O., and Schwikowski, B.~Siegel, A.~F. (2002),
  \enquote{Discovering regulatory and signalling circuits in molecular
  interaction networks,} \textit{Bioinformatics}, 18, Suppl 1:S233--40.

\bibitem[{Jaks et~al.(2001)Jaks, Jõers, Kristjuhan, and Maimets}]{jaks}
Jaks, V., Jõers, A., Kristjuhan, A., and Maimets, T. (2001), \enquote{p53
  protein accumulation in addition to the transactivation activity is required
  for p53-dependent cell cycle arrest after treatment of cells with
  camptothecin,} \textit{Oncogene}, 20, 1212--1219.

\bibitem[{Jiang and Gentleman(2007)}]{Jiang}
Jiang, Z. and Gentleman, R. (2007), \enquote{Extensions to gene set
  enrichment,} \textit{Bioinformatics}, 23, 306--313.

\bibitem[{Kanehisa and Goto(2000)}]{kegg3}
Kanehisa, M. and Goto, S. (2000), \enquote{{KEGG}: {K}yoto Encyclopedia of
  Genes and Genomes,} \textit{Nucleic Acids Res.}, 28, 27--30.

\bibitem[{Kanehisa et~al.(2006)}]{kegg2}
Kanehisa, M. et~al. (2006), \enquote{From genomics to chemical genomics: new
  developments in {KEGG},} \textit{Nucleic Acids Res.}, 34, 354--357.

\bibitem[{Karpinich et~al.(2002)Karpinich, Tafani, Rothman, Russo, and
  Farber}]{karpinich}
Karpinich, N.~O., Tafani, M., Rothman, R.~J., Russo, M.~A., and Farber, J.~L.
  (2002), \enquote{The course of etoposide-induced apoptosis from damage to
  {DNA} and p53 activation to mitochondrial release of cytochrome c,} \textit{J
  Biol Chem}, 277, 16547--16552.

\bibitem[{Khatri and Draghici(2005)}]{Khatri}
Khatri, P. and Draghici, S. (2005), \enquote{Ontological analysis of gene
  expression data: current tools, limitations, and open problems,}
  \textit{Bioinformatics}, 21, 3587--3595.

\bibitem[{Krause et~al.(2002)Krause, Wasner, Reinhard, et~al.}]{krause}
Krause, K., Wasner, M., Reinhard, W., et~al. (2002), \enquote{The tumour
  suppressor protein p53 can repress transcription of cyclin {B},}
  \textit{Nucl. Acids Res.}, 22, 4410--4441.

\bibitem[{Kurz et~al.(2004)Kurz, Douglas, and Lees-Miller}]{kurz}
Kurz, E.~U., Douglas, P., and Lees-Miller, S.~P. (2004), \enquote{Doxorubicin
  Activates ATM-dependent Phosphorylation of Multiple Downstream Targets in
  Part through the Generation of Reactive Oxygen Species,} \textit{J Biol
  Chem.}, 279, 53272--53281.

\bibitem[{Liao et~al.(2003)Liao, Boscolo, Yang, et~al.}]{liao_2003}
Liao, J., Boscolo, R., Yang, Y., et~al. (2003), \enquote{Network component
  analysis: reconstruction of regulatory signals in biological systems,}
  \textit{Proceedings of the National Academy of Sciences}, 100, 15522--15527.

\bibitem[{Lindley and Smith(1972)}]{lindley}
Lindley, D.~V. and Smith, A.~F. (1972), \enquote{Bayes estimates for the linear
  model,} \textit{Journal of the Royal Statistical Society. Series B
  (Methodological)}, 1--41.

\bibitem[{Ling and Wei-Guo(2006)}]{ling_review}
Ling, B. and Wei-Guo, Z. (2006), \enquote{p53: Structure, Function and
  Therapeutic Applications,} \textit{J Cancer Mol.}, 2, 141--153.

\bibitem[{Ling et~al.(1996)Ling, el~Naggar, Priebe, and Perez-Soler}]{ling}
Ling, Y.~H., el~Naggar, A.~K., Priebe, W., and Perez-Soler, R. (1996),
  \enquote{Cell cycle-dependent cytotoxicity, {G2/M} phase arrest, and
  disruption of p34cdc2/cyclin {B1} activity induced by doxorubicin in
  synchronized {P388} cells,} \textit{Mol Pharmacol.}, 49, 832--841.

\bibitem[{Liu et~al.(2010)}]{gnea}
Liu, M. et~al. (2010), \enquote{Network-Based Analysis of Type 2 Diabetes,}
  \textit{PLoS Genet.}, 3, e96.

\bibitem[{Lucas et~al.(2006)Lucas, Carvalho, Wang, Bild, Nevins, and
  West}]{west_2006}
Lucas, J., Carvalho, C.~M., Wang, Q., Bild, A., Nevins, J.~R., and West, M.
  (2006), \enquote{Sparse statistical modelling in gene expression genomics,}
  \textit{Bayesian Inference for Gene Expression and Proteomics}, 14,
  155–--176.

\bibitem[{Lucas et~al.(2010)Lucas, Kung, and Chi}]{lucas}
Lucas, J.~E., Kung, H.-N., and Chi, J.-T. (2010), \enquote{Latent Factor
  Analysis to Discover Pathway-Associated Putative Segmental Aneuploidies in
  Human Cancers.} \textit{PLoS Comput. Biol.}, 6, e1000920.

\bibitem[{M(1995)}]{thomasz}
M, T. (1995), \enquote{Mitomycin C: small, fast and deadly (but very
  selective),} \textit{Chemical Biology}, 2, 575--579.

\bibitem[{Ma and Zhao(2012)}]{ma}
Ma, H. and Zhao, H. (2012), \enquote{iFad: an integrative factor analysis model
  for drug-pathway association inference.} \textit{Bioinformatics}, 28,
  1911--1918.

\bibitem[{Maliga et~al.(2002)Maliga, Kapoor, and J.}]{monastrol}
Maliga, Z., Kapoor, T.~M., and J., M.~T. (2002), \enquote{Evidence that
  monastrol is an allosteric inhibitor of the mitotic kinesin Eg5,}
  \textit{Chemical Biology}, 9, 989--996.

\bibitem[{Malik et~al.(2006)Malik, Nitiss, Enriques-Rios, and Nitiss}]{malik}
Malik, M., Nitiss, K.~C., Enriques-Rios, V., and Nitiss, J.~L. (2006),
  \enquote{Roles of nonhomlogous end-joinng pathways in surviving topoisomerase
  II-mediated DNA damage,} \textit{Mol Cancer Ther}, 5, 1405.

\bibitem[{Marbach et~al.(2012)Marbach, C., Küffner, N.M., Prill, Camacho,
  Allison, Consortium, Kellis, Collins, and Stolovitzky}]{dream5}
Marbach, D., C., C.~J., Küffner, R., N.M., V., Prill, R., Camacho, D.,
  Allison, K., Consortium, T.~D., Kellis, M., Collins, J.~J., and Stolovitzky,
  G. (2012), \enquote{Wisdom of crowds for robust gene network inference,}
  \textit{Nature Methods}, 9, 796–--804.

\bibitem[{Nakada et~al.(2006)Nakada, Katsuki, Imoto, Yokoyama, Nagasawa,
  Inazawa, and Mizutani}]{nakada}
Nakada, S., Katsuki, Y., Imoto, I., Yokoyama, T., Nagasawa, M., Inazawa, J.,
  and Mizutani, S. (2006), \enquote{Early {G2/M} checkpoint failure as a
  molecular mechanism underlying etoposide-induced chromosomal aberrations,}
  \textit{J Clin Invest}, 116, 80--89.

\bibitem[{Nam and Kim(2008)}]{Nam}
Nam, D. and Kim, S.~Y. (2008), \enquote{Gene-set approach for expression
  pattern analysis,} \textit{Brief Bioinform}, 9, 189--197.

\bibitem[{Neckers et~al.(1999)Neckers, Schulte, and Mimnaugh}]{neckers}
Neckers, L., Schulte, T.~W., and Mimnaugh, E. (1999), \enquote{Geldanamycin as
  a potential anti-cancer agent: its molecular target and biochemical
  activity,} \textit{Invest New Drugs}, 17, 361--373.

\bibitem[{Obrig et~al.(1971)Obrig, Culp, McKeehan, and Hardesty}]{chx}
Obrig, T.~G., Culp, W.~J., McKeehan, W.~L., and Hardesty, B. (1971),
  \enquote{The mechanism by which cycloheximide and related glutarimide
  antibiotics inhibit peptide synthesis on reticulocyte ribosomes,}
  \textit{Journal of Biological Chemistry}, 246, 174--181.

\bibitem[{Pham et~al.(2011)Pham, Christadore, Schaus, and Kolaczyk}]{lpia}
Pham, L., Christadore, L., Schaus, S., and Kolaczyk, E.~D. (2011),
  \enquote{Network-based prediction for sources of transcriptional
  dysregulation via latent pathway identification analysis,} \textit{Proc Nat
  Acad Sci.}, 108, 13347–--13352.

\bibitem[{Pommier et~al.(2012)Pommier, Leo, Zhang, and Marchand}]{pommier}
Pommier, Y., Leo, E., Zhang, H., and Marchand, C. (2012), \enquote{DNA
  topoisomerases and their poisoning by anticancer and antibacterial drugs,}
  \textit{Chemical Biology}, 17, 421--433.

\bibitem[{Prill et~al.(2011)Prill, Saez-Rodriguez, Alexopoulos, Sorger, and
  Stolovitzky}]{dream4}
Prill, R.~J., Saez-Rodriguez, J., Alexopoulos, L.~G., Sorger, P.~K., and
  Stolovitzky, G. (2011), \enquote{Crowdsourcing Network Inference: The
  {DREAM4} Predictive Signaling Network Challenge,} \textit{Science}, 4, mr7.

\bibitem[{Rao et~al.(2012)Rao, Kurkjian, and Yamada}]{vincristine}
Rao, C.~V., Kurkjian, C.~D., and Yamada, H.~Y. (2012),
  \enquote{Mitosis-targeting natural products for cancer prevention and
  therapy,} \textit{Curr Drug Targets}, 13, 1820--1830.

\bibitem[{Reguly et~al.(2006)Reguly, Breitkreutz, Boucher,
  et~al.}]{reguly_2006}
Reguly, T., Breitkreutz, A., Boucher, L., et~al. (2006), \enquote{Comprehensive
  curation and analysis of global interaction networks in Saccharomyces
  cerevisiae,} \textit{Journal of Biology}, 5, 11.

\bibitem[{Rivals et~al.(2007)Rivals, Personnaz, Taing, et~al.}]{Rivals}
Rivals, I., Personnaz, L., Taing, L., et~al. (2007), \enquote{Enrichment or
  depletion of a {GO} category within a class of genes: which test?}
  \textit{Bioinformatics}, 23, 401--407.

\bibitem[{Roguev et~al.(2008)Roguev, Bandyopadhyay, et~al.}]{ideker_2008}
Roguev, A., Bandyopadhyay, S., et~al. (2008), \enquote{Conservation and
  rewiring of functional modules revealed by an epistasis map in fission
  yeast,} \textit{Science}, 322, 405--410.

\bibitem[{Storey and Tibshirani(2003)}]{Storey}
Storey, J.~D. and Tibshirani, R. (2003), \enquote{Statistical significance for
  genome-wide studies,} \textit{Proc. Natl. Acad. Sci.}, 100, 9440.

\bibitem[{Subramanian et~al.(2005)}]{gsea}
Subramanian, A. et~al. (2005), \enquote{Gene set enrichment analysis: A
  knowledge-based approach for interpreting genome-wised expression profiles,}
  \textit{Proc Natl Acad Sci USA}, 102, 15545--15550.

\bibitem[{Terstappen et~al.(2007)Terstappen, Schl{\"u}pen, Raggiaschi, and
  Gaviraghi}]{terstappen}
Terstappen, G.~C., Schl{\"u}pen, C., Raggiaschi, R., and Gaviraghi, G. (2007),
  \enquote{Target deconvolution strategies in drug discovery,} \textit{Nature
  Reviews}, 6, 891--903.

\bibitem[{{The Gene Ontology Consortium}(2000)}]{go}
{The Gene Ontology Consortium} (2000), \enquote{Gene Ontology: tool for the
  unification of biology,} \textit{Nature Genet.}, 25, 25--29.

\bibitem[{van Dyk and Park(2008)}]{dyk}
van Dyk, D.~A. and Park, T. (2008), \enquote{Partially collapsed {G}ibbs
  samplers: Theory and methods,} \textit{J. Amer. Stat. Soc}, 482, 790--796.

\bibitem[{Vanhaecke et~al.(2004)Vanhaecke, Papeleu, Elaut, and
  Rogiers}]{trichostatin}
Vanhaecke, T., Papeleu, P., Elaut, G., and Rogiers, V. (2004),
  \enquote{Trichostatin A-like hydroxamate histone deacetylase inhibitors as
  therapeutic agents: toxicological point of view,} \textit{Curr Med Chem}, 11,
  1629--1643.

\bibitem[{Verweij and Pinedo(1990)}]{verweij}
Verweij, J. and Pinedo, H.~M. (1990), \enquote{Mitomycin {C}: mechanism of
  action, usefulness and limitations,} \textit{Anticancer Drugs}, 1, 5--13.

\bibitem[{Vogelstein and Kinzler(2004)}]{vogelstein}
Vogelstein, B. and Kinzler, K.~W. (2004), \enquote{Cancer genes and the
  pathways they control,} \textit{Nature Medicine}, 10, 789--799.

\bibitem[{von Mering et~al.(2002)von Mering, Krause, Snel, Cornell,
  et~al.}]{von_2002}
von Mering, C., Krause, R., Snel, B., Cornell, M., et~al. (2002),
  \enquote{Comparative assessment of large-scale data sets of protein-protein
  interactions,} \textit{Nature}, 417, 6887.

\bibitem[{Wang et~al.(2004)Wang, Konorev, Kotamraju, Joseph, Kalivendi, and
  Kalyanaraman}]{wang}
Wang, S., Konorev, E., Kotamraju, S., Joseph, J., Kalivendi, S., and
  Kalyanaraman, B. (2004), \enquote{Doxorubicin Induces Apoptosis in Normal and
  Tumor Cells via Distinctly Different Mechanisms,} \textit{J Biol Chem.}, 279,
  25535--25543.

\bibitem[{Wei and Pan(2008)}]{wei}
Wei, P. and Pan, W. (2008), \enquote{Incorporating gene networks into
  statistical tests for genomic data via a spatially correlated mixture model,}
  \textit{Bioinformatics}, 24, 404--411.

\bibitem[{Yamanishi et~al.(2008)Yamanishi, Araki, Gutteridge, Honda, and
  Kanehisa}]{yamanishi}
Yamanishi, Y., Araki, M., Gutteridge, A., Honda, W., and Kanehisa, M. (2008),
  \enquote{Prediction of drug-target interaction networks from the integration
  of chemical and genomic spaces,} \textit{Bioinformatics}, 24, 232--240.

\bibitem[{Zhou et~al.(2002)Zhou, Gu, Li, Zhu, Woods, and Findley}]{zhou}
Zhou, M., Gu, L., Li, F., Zhu, Y., Woods, W.~G., and Findley, H.~W. (2002),
  \enquote{{DNA} damage induces a novel p53-survivin signaling pathway
  regulating cell cycle and apoptosis in acute lymphoblastic leukemia cells,}
  \textit{Mol Pharmacol}, 303, 124--31.

\end{thebibliography}
} \fi
\end{document}